\documentclass[12pt]{iopart}

\usepackage{iopams}  
\usepackage[dvipsnames]{xcolor}
\usepackage{hyperref}
\hypersetup{hidelinks}
\usepackage{graphicx}
\usepackage{relsize}
\usepackage[normalem]{ulem}

\begin{document}

\title{Complex Band Structure and Bound States in the Continuum: A Unified Theoretical Framework}

\author{Jie Liu$^{1, 4}$, Ziyun Peng$^{1, 4}$, Qianju Song$^{1,2,*}$, Ang Chen$^{1}$, Liping Yang$^{1}$, Chunxiong Zheng$^{3}$, and  Dezhuan Han$^{1,*}$}

\address{$^{1}$School of Physics, Chongqing University, Chongqing 401331, China}
\address{$^{2}$School of Mathematics and Physics, Southwest University of Science and Technology, Mianyang 621010, China}
\address{$^{3}$Department of Mathematical Sciences, Tsinghua University, Beijing 100084, China}
\address{$^{4}$Authors contributed equally}
\ead{qjsong@swust.edu.cn, dzhan@cqu.edu.cn}

\vspace{10pt}

\begin{abstract}

Band structure analysis is central to understanding wave propagation in periodic media; however, it becomes challenging in open systems owing to energy leakage. Photonic crystal (PhC) slabs exemplify such systems, featuring periodicity in the $x$-$y$ plane and finite extent in the $z$-direction, and supporting diverse guided-mode resonances whose interactions give rise to phenomena such as bound states in the continuum (BICs), exceptional points (EPs), and circular polarisation states. Although numerical simulations can reveal these effects, effective non-Hermitian Hamiltonians are often employed to elucidate the underlying physical mechanisms. This approach, however, relies on manually selected resonant modes and may suffer from basis incompleteness. Here, a systematic first-principles approach is presented to derive the complex band structure. The minimal channels in the scattering matrix, either open or closed, are determined by the number of propagating bulk Bloch waves. The interactions between these waves fully reveal the complex band structure. For instance, two Bloch waves predict the leading-order imaginary frequency $\omega''$ and identify accidental BICs, each associated with a dual Fabry--Pérot mode, whereas three waves reveal robust Friedrich--Wintgen and symmetry-protected BICs together with the associated linewidth behaviours. Orthogonally polarised waves are further incorporated to characterise far-field polarisation and EPs. When extended to a two-dimensional periodic structure, this framework accurately predicts $\omega''$, encompasses all known BICs, and tracks their evolution with system parameters. Overall, this first-principles approach provides a unified foundation for studying complex band structure and facilitates the exploration of light confinement in periodic media.

\end{abstract}

%
\vspace{2pc}
\noindent{\it Keywords}: Complex band structure, Fabry--Pérot resonance, Guided-mode resonance, Bound state in the continuum, Exceptional point
%
%
%
%

\section{Introduction} 
\label{sec:introduct}

The periodicity of a periodic structure gives rise to discrete translational symmetry, resulting in a dispersion relation that manifests as an energy band structure. In addition to the form of the periodic potential, the energy band structure depends on the type of wave equation governing the system. Here, electromagnetic (EM) waves are considered. If the potential is uniform or periodic in all relevant directions, the physical system is Hermitian and exhibits a continuum, or multiple continua, of extended states with real energy eigenvalues. In contrast, structures such as photonic crystal (PhC) slabs~\cite{joan08, sako05}, which are periodic in the $x$-$y$ plane, have finite thickness in the $z$-direction, and are embedded in a background medium, constitute open systems owing to their interaction with the surrounding environment. Outside the light cone, the operator governing time evolution remains Hermitian, and well-defined energy bands can be obtained. However, modes within the slab cannot be perfectly confined inside the light cone because of the presence of a continuous spectrum in the background and the fact that interfaces in the $z$-direction generally do not act as perfect mirrors for dielectric media. Instead, they become leaky modes or resonant modes \cite{quar18}, dubbed guided-mode resonances, which possess finite \textit{Q} factors and give rise to complex band structure. When excited, these modes appear as peaks in the scattering spectrum, whose linewidths can vary sharply.

These guided-mode resonances have been studied extensively, both theoretically and experimentally. Bound states in the continuum (BICs) constitute a particular class of discrete states within guided-mode resonances, characterised by purely real energy eigenvalues and zero linewidth in the spectrum~\cite{neum29, fried85, hsu13}. More recently, a technique known as Brillouin zone (BZ) folding, induced by a period-doubling perturbation, has enabled guided modes originally located outside the light cone to fold into the first BZ and couple to the radiation continuum, thereby giving rise to extremely high-\textit{Q} modes~\cite{zeng2015tunable, overvig2018dimerized, wang2023brillouin, Zhou2025tailoring}. By exploiting the strong field confinement of these high-\textit{Q} modes, enhanced nonlinear optical effects~\cite{carletti2018giant, liu2019high}, strengthened light--matter interactions \cite{maggiolini2023strongly, weber2023intrinsic}, and lasing action \cite{kodigala2017lasing, noda2023high} can be realised. Furthermore, the far-field radiation of these modes exhibits notable physical properties, with circular polarisation states being of particular interest owing to their connection to the chiral response of resonant metasurfaces~\cite{gorkunov2020metasurfaces, overvig2021chiral, liu2019circularly, chen2023observation, kuhner2023unlocking}. By combining near- and far-field properties, polarisation singularities such as BICs and band singularities such as exceptional points (EPs)~\cite{bend98, rute10, feng13} can be utilised simultaneously~\cite{wang2025photoswitchable, deng2022extreme}. Moreover, the topological vortices associated with BICs in momentum space can be exploited to generate and manipulate optical vortex beams~\cite{wang2020generating, huang2020ultrafast}, even in the presence of real-space disorder~\cite{qin2025disorder}.

Resonant modes within the light cone can be analysed theoretically using a framework analogous to that employed for eigenmodes outside the light cone through the introduction of an effective non-Hermitian Hamiltonian. Effective non-Hermitian Hamiltonians have long been studied in nuclear, atomic, and other quantum systems \cite{fesh58, rott09, sadr21}. In the context of optical waves, a closely related formulation is provided by coupled-mode theory~\cite{haus84, mano99, fan02}, which incorporates equations describing both closed and open channels. The undetermined parameters in this framework are typically obtained by fitting resonance peaks, from which the complex energy band structure can be derived.

Although the non-Hermitian Hamiltonian is effective, it is not derived from first principles and is primarily employed to interpret numerical or experimental results. For a complex energy band, a rigorous definition should instead begin with the poles of the scattering matrix. Simple poles in the complex energy plane correspond one-to-one with scattering peaks. The real part, $\omega'$, determines the peak position, whereas the imaginary part, $\omega''$, determines the peak width. A nonzero imaginary component represents the energy decay rate of quasi-normal modes, and the \textit{Q} factor can be defined as $\omega'/2\omega''$. The poles of the scattering matrix can be computed using various methods, including plane-wave \cite{liu2012s4, yang14}, Bloch-wave \cite{lala06, hu23}, waveguide-mode \cite{chang12}, and finite element formulations, all of which fundamentally require solving the problem in a sufficiently large Hilbert space. By contrast, effective Hamiltonians are typically low-dimensional. This disparity motivates the need to reduce the scattering matrix derived from first principles to a minimal Hamiltonian, or equivalently, to the lowest-dimensional Hilbert space capable of capturing the essential physics.

In this study, the minimal Hilbert space required to estimate the poles of the scattering matrix is identified from first principles. The physical picture is straightforward: the process is primarily governed by bulk Bloch waves propagating in the $z$-direction within the slab. One or multiple such waves may be present, and in the background medium surrounding the slab, an equal number of diffraction orders must be considered to satisfy the boundary conditions and ultimately form resonant modes. The waves in the background medium, corresponding to different diffraction orders with either real or imaginary $k_z$ in the $z$-direction, act as open or closed channels in the scattering-matrix formalism. Therefore, the number of scattering channels is equal to the dimension of Hilbert space or the number of bulk Bloch waves.

Using this framework, the complex energy bands are rigorously derived using perturbation theory, yielding a general expression for the imaginary parts in addition to that for the real parts. Owing to its negative semidefiniteness, the imaginary component depends on the asymptotic perturbation parameter in a $\delta^2$ manner, with the proportionality coefficient $C(\textbf{k}_{||})$ (or $C(q)$ for a specified direction) determined solely by the lattice type, the slab thickness in the $z$-direction, and the Bloch wavevector, analogous to the role of the structure factor. It is found that $C(q)$ possesses well-defined zeros corresponding to BICs. Thus, BICs arise from the breaking of continuous translational symmetry into discrete translational symmetry and can be treated as fixed points that are robust against perturbations of the periodic potential. In addition to accidental BICs arising from the interaction of two Bloch waves, interactions among three Bloch waves are shown to give rise to conventional Friedrich--Wintgen BICs \cite{fried85} and symmetry-protected BICs \cite{fan02}. The existence of Friedrich--Wintgen BICs is analytically demonstrated and identified in the vicinity of energy-band crossing points.

This theory also employs first principles to explain the evolution of resonance peak widths in parameter space by expanding $C(q) \approx \tilde{C}\,(q - q_{\mathrm{BIC}})^2$ \cite{yuan2017strong}. This result is consistent with previous experimental and numerical observations, which show that some narrow peaks remain well preserved in momentum space, whereas others broaden significantly. Far-field polarisation states and polarisation singularities can be analysed by incorporating polarisation degrees of freedom. Furthermore, EPs emerge when interactions between Bloch waves of orthogonal polarisation are considered away from high-symmetry lines. Finally, the theory is generic and equally applicable to two-dimensional (2D) PhC slabs. In this study, guided-mode resonances in 2D PhC slabs, associated with bands folded along different directions, were systematically analysed, encompassing all possible cases of accidental BICs and Friedrich--Wintgen BIC configurations.

\section{Methods} 
\label{sec:methods}

\subsection{Formalism based on scattering matrix} \label{sec:S-matrix}
\label{subsec:smatrix}

In quantum mechanics, resonant states arise in a one-dimensional (1D) potential well, with corresponding energies located in the fourth quadrant of the complex plane, where $\mathrm{Re}(E) > 0$ and $\mathrm{Im}(E) < 0$, ensuring that the physical solutions do not diverge as $t \rightarrow \infty$ \cite{newt82}. These states are analogous to Fabry--Pérot (FP) resonances in optical films. A commonly used definition of FP resonances is based on unit transmission, $T = 1$, which describes wave propagation within the film with an optical path length equal to half-integer or integer multiples of the wavelength. Below the light cone, different branches of waveguide modes correspond to bound states with $E < 0$ in quantum mechanics. When periodicity is introduced into films—such as in PhC slabs that are finite in the $z$-direction and periodic in the $x$-$y$ plane---the waveguide modes fold into the first BZ, enter the light-cone regime, and become guided-mode resonances. Consequently, unlike uniform planar waveguides, both guided-mode and FP resonances coexist within the light-cone regime, and their interplay must be carefully considered \cite{hu2022global}.

The dispersion of guided-mode resonances is determined by the poles of the scattering matrix, i.e. $\mathrm{det}(S^{-1}) = 0$. As a class of quasi-normal modes, guided-mode resonances can significantly enhance the near field and exhibit long tails in the far-field radiation. The near-field wavefunction can be analysed by decomposing it into \textit{bulk} Bloch waves \cite{hu23,dai2018topologically}, defined here as Bloch waves that are not confined in the $z$-direction but that account for periodicity in the $x$-$y$ plane. The far-field radiation of guided-mode resonances, whose propagation directions are determined by the frequency and in-plane wavevector, gives rise to a polarisation field in momentum space \cite{zhen2014topological,zhang2018observation,doeleman2018experimental}. The singularities of this polarisation field are related to the topology of the two-dimensional BZ. The polarisation field is defined for a fixed diffraction order and is typically considered only for propagating, rather than evanescent, waves \cite{jiang2023general,berry01}. The case in which only a single diffraction order produces a propagating far-field wave is therefore considered. In this context, BICs correspond to a special class of singularities characterised by integer topological charges of the polarisation field. These integer charges, however, are not intrinsically stable and require protection through specific symmetries \cite{zhen2014topological}. When such symmetries are broken, the charges can split into pairs of half-integer charges, known as circular polarisation points \cite{liu2019circularly, yin2020observation, chen2021evolution}.

For simplicity, without loss of generality, we first consider scalar waves, such as transverse-electric (TE) polarised electromagnetic waves in a PhC slab of thickness $h$ in the $z$-direction, period $a$ in the $x$-direction, and uniformity in the $y$-direction. Noted that throughout this work, TE refers to modes with the electric field parallel to the $x$-$y$ plane, while TM refers to modes with the electric field perpendicular to the $x$-$y$ plane. This classification is consistent with the convention for PhC slabs~\cite{joan08}. For incident waves with wavevector $q$, the wavefunction (electric field $E$) in the three regions can be expressed as:
\begin{equation} 
\Psi(x,z) = \left\{ 
\begin{array}{lll} 
&\sum\limits_{m} \left[ a^+_m e^{i k^{\mathrm{b}}_{zm}z} + a^-_m e^{-i k^{\mathrm{b}}_{zm}z} \right] e^{i(q+mG)x}, & z < -h/2, \\ 
&\sum\limits_{n}\left[c_n e^{ik_{zn}z} + d_n e^{-ik_{zn}z} \right] \psi_n(x), & z \in [-h/2,h/2], \\ 
&\sum\limits_{l} \left[ b^-_l e^{i k^{\mathrm{b}}_{zl}z} + b^+_l e^{-i k^{\mathrm{b}}_{zl}z} \right] e^{i(q+lG)x}, & z > h/2, 
\end{array} 
\right. 
\label{eq:wavefunc-slab}
\end{equation}

\noindent where $G = 2\pi/a$ denotes a reciprocal primitive vector, and $\psi_n$ represents the wavefunction of the Bloch state with wavevector $q$ in the $n$-th band. The $z$-components of the wavevectors are denoted by $k_z^\mathrm{b}$ and $k_z$ for the background and the slab, respectively. Here, we assume $k_y = 0$, and the case $k_y \neq 0$ is analysed later. Figure 1 illustrates the scattering problem considered for a PhC slab, either one- or two-dimensional. If we define $\Psi^+ = \{\dots, a^+_m, \dots;\dots, b^+_l, \dots \}^\mathrm{T}$ and $\Psi^- = \{ \dots, a^-_m, \dots;\dots, b^-_l, \dots \}^\mathrm{T}$, the scattering matrix $S$ relates the incoming and outgoing waves as:
\begin{equation}
    \Psi^- = S \Psi^+.
    \label{eq:S-matrix}
\end{equation}
The matrix $S$ is unitary if energy flux is conserved, implying that it does not possess poles on the real-$\omega$ axis. However, poles may exist in the complex plane and can be determined numerically.

\begin{figure}[h]
    \centering    
    \includegraphics[width=0.66\textwidth]{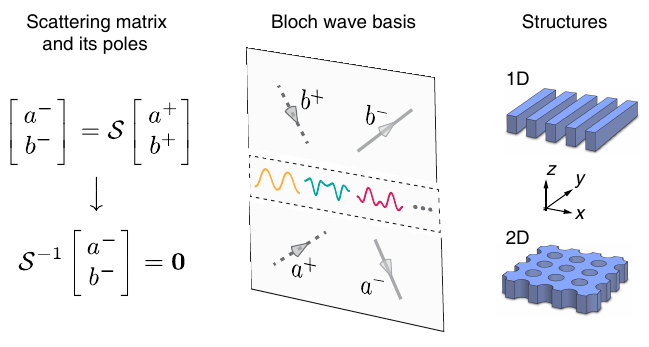}
    \caption{Scattering matrix and its poles in the complex-$\omega$ plane. The scattering matrix relates the incoming ($a^+$, $b^+$) and outgoing ($a^-$, $b^-$) waves. Inside the periodic structure, whether one-dimensional or two-dimensional, a Bloch-wave basis is adopted to describe the electromagnetic fields.
    }
    \label{fig:S-matrix}
\end{figure}

\subsection{Formalism of perturbation theory}
\label{subsec:perturbtheory}

To capture the physical essence and obtain a clear picture, we perform an asymptotic analysis on the poles of the $S$-matrix. In non-magnetic dielectric media, only the dielectric function $\epsilon$ varies spatially. The expressions for the electric and magnetic fields are given as:
\begin{equation}
     \nabla \times \nabla \times \mathbf{E} - \epsilon(\mathbf{r}) k_0^2 \mathbf{E} = 0, 
    \label{eq:masterEq-E}
\end{equation}
\begin{equation}
     \nabla \times \frac{1}{\epsilon(\mathbf{r})} \nabla \times \mathbf{H} - k_0^2 \mathbf{H} = 0, \label{eq:masterEq-H}
\end{equation}
in which $k_0=\omega/c$ is the wavevector in the free space. For a TE-polarised wave with wavevector $\mathbf{k}=(q,0,k_z)$, $\mathbf{E}$ reduces to a scalar and the equation is simplified to a Helmholtz equation in the inhomogeneous medium, as follows:
\begin{equation}
\nabla^2 \Psi(\textbf{r})+k_{0}^{2} \epsilon(\textbf{r})\Psi(\textbf{r})=0.\label{eq:ModifiedHelmholtz}
\end{equation}

\noindent For the 1D case, the perturbed potential can be written as: 
\begin{equation}
\epsilon(\textbf{r})=
\left\{
\begin{array}{ll}
     \epsilon_{\mathrm{b}} \quad & \mathrm{for} \ |z| > h/2, \\
     \bar{\epsilon} + \varepsilon(x)\delta \quad & \mathrm{for} \ |z| \leq h/2.
\end{array} \label{eq:epperiodic}
\right.
\end{equation}
In the above, $\delta \ll 1$ is singled out as a small asymptotic parameter, and the periodic modulation $\varepsilon(x)=\varepsilon(x+a)$ is  O$(1)$. The average permittivity $\bar{\epsilon}$ inside the PhC is explicitly treated separately. We will demonstrate later that this simplest case can capture the essential physics for the estimation of the imaginary part $\omega''$. At $z=0$, the homogeneous Dirichlet boundary condition, $\Psi(x,0)=0$, is applied to the odd TE modes, as they exhibit nodes at $z=0$. For the even modes, the Neumann boundary condition, $\partial_z \Psi(x,0)=0$, is adopted, as they have extrema at $z=0$.

\subsection{Two-step approach to perturbation theory}
\label{subsec:2-stepperturb}

The perturbation theory can be divided into two steps. In the first step, we solve the \textit{bulk} Bloch waves without accounting for the boundaries at $z = \pm h/2$. We choose a basis for the unperturbed system, which is a homogeneous medium with a dielectric constant $\bar{\epsilon}$, given by $e^{i q_n x}$ with $q_n = q + n G$. Hereinafter, we also represent them using the Dirac ket notation $|n\rangle$. Then, we consider the following perturbed eigenvalue problem:

\begin{equation}
    \partial_{xx} \psi(x) + k_0^2 \bigl( \bar{\epsilon} + \varepsilon(x)\delta \bigr) \psi(x) = \lambda \psi(x),
    \label{eq:laplacian1dpert}
\end{equation}

\noindent where $\psi(x) \equiv \psi^{(\delta)}(x)$, and we omit the superscript $\delta$ below. It is evident that $|\psi_n \rangle \rightarrow |n \rangle$ as the asymptotic parameter $\delta \rightarrow 0$. For $\delta \neq 0$, applying the standard perturbation theory yields the following results:
\begin{equation}
        \lambda_n = \bar{\epsilon} k_0^2 - q_n^2 + \mathrm{O}(\delta^2), 
\label{eq:eigenvaluepertur}
\end{equation}
\begin{equation}
    |\psi_n\rangle = |n \rangle + \Big( \sum_{m \neq n } \frac{k_0^2 \langle m|\varepsilon(x)|n \rangle}{q_m^2-q_n^2} |m \rangle \Big) \delta + \mathrm{O}(\delta^2).
\label{eq:eigenstatepertur}
\end{equation}
\noindent From the above, it can be seen that the correction to the eigenvalue $\lambda \ \bigl(\equiv k_z^2 \bigr) $ is $\mathrm{O}(\delta^2)$, and the same holds for $k_z$. This is because its first-order correction arises from the variation in $\bar{\epsilon}$. Because we keep $\bar{\epsilon}$ invariant and ensure that the unit cell average satisfies $\langle\varepsilon(x)\rangle=0$ in the equation (\ref{eq:epperiodic}), the first-order correction vanishes. We also note that the eigenvalue $\lambda > 0$ corresponds to the propagating Bloch waves of interest in this study, whereas $\lambda < 0$ corresponds to the evanescent Bloch waves that can be ignored in the context of perturbation. We further introduce a matrix $\textbf{U}=(u_{nm})$ defined as:
\begin{equation*}
    u_{nm} = \left\{
    \begin{array}{ll}
    \langle m|\varepsilon(x)|n \rangle k_0^2 \big/ (q_m^2-q_n^2) \quad & \mathrm{for} \ m \neq n, \\
    0 \quad & \mathrm{for} \ m=n. 
    \end{array}
    \right.
\end{equation*}
\noindent We can rewrite equation~(\ref{eq:eigenstatepertur}) as:
\begin{equation}
    |\psi_n\rangle = |n\rangle + \sum_{m} u_{nm} |m\rangle \delta + \mathrm{O}(\delta^2),
\label{eq:eigenstatepertur-1}
\end{equation}
or its equivalent in the vector form as:
\begin{equation*}
    \boldsymbol{\psi} = \big( \textbf{I} + \textbf{U} \delta \big) \boldsymbol{\psi}^0 + \mathrm{O}( \delta^2),
\end{equation*}
where $\boldsymbol{\psi}^0 = (\cdot\cdot\cdot,|n\rangle,\cdot\cdot\cdot)^\mathrm{T}$. Because $\textbf{U}$ is anti-Hermitian, i.e. $\textbf{U}^{\dag} = -\textbf{U}$, it can be directly verified that the wave function $\psi$ is normalised up to $\mathrm{O}(\delta^2)$.

For the second step of the perturbation, we adopt the above \textit{bulk} Bloch states presented in equation~(\ref{eq:eigenstatepertur}) and apply them to equation~(\ref{eq:wavefunc-slab}), specifically replacing the wave function within the slab region $z \in [-h/2,h/2]$. To determine the boundary conditions at infinity, we need to fix $\{\dots, a^+_m, \dots \}^\mathrm{T}$ for $z<-h/2$, and $\{ \dots,b^+_l,\dots\}^\mathrm{T}$ for $z>h/2$, which correspond to the incident waves $\Psi^+$. A single incident wave is usually considered for a typical scattering problem; for example, we can set a wave incident from the region $z<-h/2$ so that $\Psi^+ = \{\dots,0,1,0,\dots;0,\dots,0\}^\mathrm{T}$. The scattered waves can then be  calculated directly as $\Psi^-=S \Psi^+$. For the non-lossy materials considered here, the energy flux is conserved and $S$-matrix is unitary. However, if the $S$-matrix has a pole, a nonzero $\Psi^-$ can be obtained even in the absence of an incoming wave ($\Psi^+ = 0$). Consequently, the poles of the $S$-matrix lie on the complex-$\omega$ plane. The estimation of their imaginary parts from the condition:
\begin{equation}
    \mathrm{det}\big[S^{-1}(\omega'-i\omega'')\big]=0,
    \label{eq:S-poles}
\end{equation}
forms the central focus of this study. Note that the convention of a negative sign in the imaginary part ensures the positive semi-definiteness of $\omega''$. 

However, the advantages of using the basis of Bloch waves have not yet been fully demonstrated in the above analysis. In fact, when employing a sufficiently large Hilbert space for numerical simulations, the choice between Bloch waves and plane waves does not make a significant difference. In the following, we show that the minimal Hilbert-space dimension is determined by the number of propagating bulk Bloch waves involved, i.e.,
\begin{equation}
    \mathrm{dim}_{\mathrm{min}}(S) = N_{\mathrm{propagating\ Bloch}},
    \label{eq:S-min-rank}
\end{equation}
which is sufficient to capture the essential physics and to achieve accurate calculations within the framework of the perturbation theory. This approach not only provides a clear physical picture but also determines the poles of the $S$-matrix in the complex-$\omega$ plane. In fact, equation~(\ref{eq:S-poles}) is a rank-deficiency condition for the $S$-matrix. By using a minimal Hilbert space, the origin of complex band structure and formation of various BIC types can be systematically analysed. As a first example, we discuss `accidental' BICs, which arise from the interaction between only two propagating Bloch waves. We then explore Friedrich--Wintgen BICs and symmetry-protected BICs, as well as the high $Q$ resonant states in their vicinity, which result from the interaction between three Bloch waves. Furthermore, we analyse the far-field radiation of resonant states and emergence of EPs. Because the polarisation degrees of freedom are included, the number of Bloch waves doubles—four Bloch waves are required for polarisation singularities and six Bloch waves for EPs. Finally, we extend the framework to 2D PhC slabs and examine all types of interactions between different energy bands. The formation and evolution of accidental and Friedrich--Wintgen BICs are analysed using the minimal set of bulk Bloch waves.

\section{Results and discussion} \label{sec:Results and discussion} 
\begin{figure}[h]
    \centering    \includegraphics[width=0.66\textwidth]{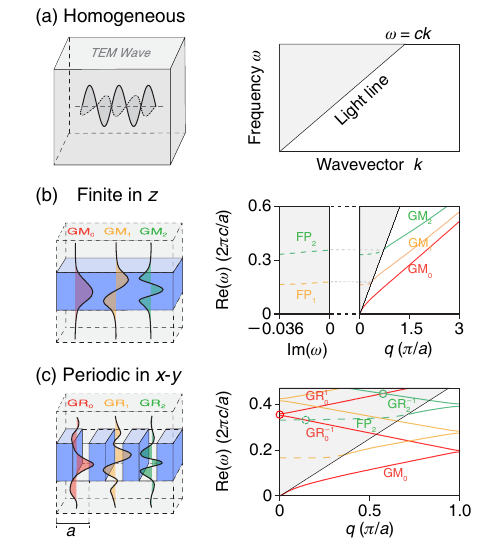}
    \caption{EM waves in different media. Left: field profiles in (a) homogeneous medium, (b) slab waveguide, and (c) PhC slab. Right: corresponding dispersion relations. (a) The region above the light line (shaded grey) supports wave propagation. (b) FP modes (complex frequencies) lie above the light line, whereas waveguide modes (real frequencies) lie below it. (c) Waveguide modes folded into the first BZ interact with the FP modes, forming guided mode resonances with complex frequencies. Parameters: $\bar\epsilon=9$, $h=a=600\,$nm; perturbation strength $\delta=0.3$.
    }
    \label{fig:free-wg-PhC} 
\end{figure}

The problem of a \textit{minimal} Hilbert space does not arise in an unperturbed homogeneous waveguide, where $\delta = 0$. Because the system possesses continuous in-plane translational symmetry, the wavevector $\mathbf{k}_{||}$ in the $x$-$y$ plane is a good quantum number and can be used to distinguish different waves. Therefore, the corresponding Hilbert space reduces to a trivial 1D case. 
In figure~\ref{fig:free-wg-PhC}(a), we illustrate that plane waves in a homogeneous medium form a continuum of states. When confinement is introduced in the $z$-direction, waveguide modes can be formed below the light line, as shown in figure~\ref{fig:free-wg-PhC}(b). 
Above the light line, these waveguide modes continuously transition into FP modes, which are a typical type of resonance and can be excited by the states in the continuum. When a perturbation is introduced, such that $\delta \neq 0$, all branches of the waveguide modes, which have no upper bound on $k_{||}$, are folded into the first BZ, forming guided-mode resonances. Here, we examine two key issues in detail. First, we analyse the poles of the $S$-matrix corresponding to the FP modes, in a manner similar to the treatment of the waveguide modes. Second, we investigate the interaction between different bands of FP modes and guided-mode resonances, as described later in multiple subsections.

For an unperturbed system, setting $\Psi^+ = \big( a^+,b^+ \big)^\mathrm{T} = 0$ reduces the $S$-matrix to a $2\times2$ form, and the condition $S^{-1} \Psi^- = 0$ leads to
\begin{equation}    
    \frac{\partial_z\Psi}{\Psi} \bigg|_{z=h/2} = i k_{z}^{\mathrm{b}}, 
    \label{eq:wgm-DtN}
\end{equation}
where $k^\mathrm{b}_{z}$ is the $z$-component of the wavevector in the background medium. The above condition, also known as the Sommerfeld radiation condition, can be reduced to the case of even modes by substituting $\Psi \rightarrow \cos(k_zh/2)$ and to the case of odd modes by substituting $\Psi \rightarrow \sin(k_zh/2)$. 
Mathematically, the left-hand and right-hand sides of equation~(\ref{eq:wgm-DtN}) correspond to the Dirichlet-to-Neumann (DtN) operators for the periodically structured slab and background medium, respectively. Physically, they represent the surface admittance when $\Psi$ denotes the electric field, or the surface impedance when $\Psi$ denotes the magnetic field.
Equation~(\ref{eq:wgm-DtN}) is, in fact, an impedance matching condition. When analysing the TE modes, it is convenient to consider $\Psi$ as the electric field, as adopted in this study.
  
The dispersion of the waveguide modes can be determined from equation~(\ref{eq:wgm-DtN}) when $q > n_{\mathrm{b}} k_0$, where $n_{\mathrm{b}} = \sqrt{\epsilon_{\mathrm{b}}}$ is the background refractive index. These modes lie below the light line, as shown in the right panel of figure~\ref{fig:free-wg-PhC}(b), and are referred to as GM$_m$ for the $m$-th mode, having $m$ nodes inside the waveguide. 
Here, the permittivities are selected as $\epsilon_{\mathrm{b}} = 1$ and $\bar{\epsilon} = 9$. 
If $q \leq n_{\mathrm{b}} k_0$, the outgoing waves are no longer evanescent but propagating as $z \rightarrow \pm \infty$. 
Equation~(\ref{eq:wgm-DtN}) can be solved only in the complex-$\omega$ plane, where the real part $\omega'$ corresponds to the dispersion of the FP modes. 
In figure~\ref{fig:free-wg-PhC}(b), the imaginary part $\omega''$ is also plotted, which represents the width of the FP modes and is inversely proportional to their finesse. 
Note that the odd TE modes exhibit a lower cutoff frequency, whereas the even modes have no cutoff frequency, as the $\mathrm{TE}_0$ mode passes through the origin of the $(q, \omega)$ plane.

Next, let us consider the perturbed eigenvalue problem for a slab with $\delta \neq 0$. The wave function $\Psi$ inside the slab can be expanded in terms of the perturbed bulk Bloch-wave functions given in equation~(\ref{eq:eigenstatepertur-1}):
\begin{equation}
    \Psi(x,z) = \sum_{n} \left( c_n e^{ik_{zn}z} + d_n e^{-ik_{zn}z} \right) \psi_n(x).
    \label{eq:wfuncslabexpand}
\end{equation}
Specifically, $\Psi$ can be re-expressed in the plane-wave basis at the interface $z = h/2$ as $\Psi(x, h/2) = \sum_{m} \tilde{u}_m |m\rangle$. A similar expansion can be performed at the opposite interface $z = -h/2$, which yields no additional information when the system possesses $\sigma_h$ mirror symmetry (i.e. reflection across the $x$-$y$ plane). For odd modes, the substitution $c_n e^{i k_{zn} z} + d_n e^{-i k_{zn} z} \rightarrow c_n \sin(k_{zn} z)$ can be applied. We define the vectors $\textbf{c} = (\cdots,c_n,\cdots)^\mathrm{T}$ and $\tilde{\textbf{u}} = (\cdots,\tilde{u}_m,\cdots)^\mathrm{T}$ for the wave function in the basis of the bulk Bloch waves and plane waves, respectively. This change in basis is represented by a matrix $\mathcal{U}$, which relates $\tilde{\textbf{u}}$ and $\textbf{c}$ in the following form:
\begin{equation}
    \tilde{\textbf{u}} = \mathcal{U} \, \mathrm{diag} \left( \sin(k_{zn}h/2) \right) \, \textbf{c}.
    \label{eq:change-of-basis}
\end{equation}
The derivative of the wave function is given by:
\begin{equation*}
\partial_z \Psi \big|_{z=h/2} = \sum_{m} \left( \mathcal{U} \, \mathrm{diag} \left(k_{zn} \cos(k_{zn}h/2)\right) \, \textbf{c} \right)_m |m\rangle.
\end{equation*}
In the above, the main objective is the analysis of the DtN operator, very similar to that performed in equation~(\ref{eq:wgm-DtN}). 
The DtN operator can be expressed as a \textit{diagonal matrix} for both the Bloch waves in the bulk PhC and plane waves in the background (including evanescent waves when $q > n_{\mathrm{b}} k_0$). By matching the boundary condition at $z = \pm h/2$, we obtain:
\begin{equation*}
    \Bigg[ \mathcal{U}\, \mathrm{diag}\left( \frac{k_{zn} \cos(k_{zn} h/2)}{\sin(k_{zn} h/2)} \right)\, \mathcal{U}^{-1} - \mathrm{diag} \left(i k_{zm}^{\mathrm{b}} \right) \Bigg]\,\tilde{\textbf{u}} = 0.
\end{equation*}
The change-of-basis matrix, $\mathcal{U}$, plays a key role here. The above equation holds true for even modes by replacing $\Psi$ from $\sin(k_{zn} z)$ with that from $\cos(k_{zn} z)$. 
Therefore, it can be written in the following compact form:
\begin{equation}
    \mathrm{det}\left[ \mathcal{U}\, (\mathrm{DtN})_{\mathrm{PhC}}\, \mathcal{U}^{-1} - \mathrm{diag} \left(i k_{zm}^{\mathrm{b}} \right) \right] = 0,
\label{eq:DtN^PhC vs DtN^b}
\end{equation}
where $(\mathrm{DtN})_{\mathrm{PhC}}$ denotes the DtN matrix in the bulk PhC. The above equation is generally valid, even in the case of a large index contrast. In the context of perturbation, matrix $\mathcal{U} \approx \textbf{I} + \textbf{U} \delta$, with the matrix $\textbf{U}$ defined in equation~(\ref{eq:eigenstatepertur-1}), and the above equation becomes:
\begin{equation}
    \mathrm{det}\left[ \big(\textbf{I} + \textbf{U} \delta\big)(\mathrm{DtN})^{\delta} \big(\textbf{I} - \textbf{U} \delta\big) - \mathrm{diag} \left(i k_{zm}^{\mathrm{b}} \right) \right] = 0,
\label{eq:DtN^delta vs DtN^b}
\end{equation}
where $(\mathrm{DtN})^{\delta}$ denotes the DtN matrix in the bulk PhC with a perturbed dielectric function when $\delta$ is finite.

\subsection{Origin of Im($\omega$): interaction between FP modes and guided-mode resonances} \label{subsec: interaction of FP & GMR}

In the above, equations~(\ref{eq:wgm-DtN}) and (\ref{eq:DtN^delta vs DtN^b}) present two limiting cases: the first one considers a 1D Hilbert space, whereas the second involves an infinite-dimensional space. Here, we focus on whether the infinite-dimensional Hilbert space in equation~(\ref{eq:DtN^delta vs DtN^b}) can be reduced to a minimal one while retaining the physical essence. 
Among all optical modes, the guided-mode resonances, as depicted in figure \ref{fig:free-wg-PhC}(c), are of particular interest. 
The mode $\mathrm{GR}_n^m$ originates from the waveguide mode GM$_{n}$, with $m$ being the band-folding index. Here, we use the period $a=600$ nm; the periodic modulation $\varepsilon(x)=1$ for $-a/2<x<a/4$, $\varepsilon(x)=-3$ for $a/4<x< a/2$, and the perturbation strength $\delta=0.3$. For the lowest band, the $\mathrm{GR}_0^0$ mode (with no folding), considering only a 1D Hilbert space and applying equation~(\ref{eq:wgm-DtN}), provides the dispersion in relation to the leading order. However, a significant challenge arises when applying this equation to the GR$_0^{-1}$ modes above the light line. The 1D Hilbert space cannot yield a nonzero imaginary part $\omega''$, as the leading order of $\omega''$ from the GM$_0$ mode, a waveguide mode with a real $\omega$, is zero. Therefore, the minimal dimension must be $\geq 2$ to obtain a nonzero $\omega''$.

The requirement that the minimal dimension $\geq 2$ applies only to the guided-mode resonances. The FP modes above the light line already possess a nonzero $\omega''$ according to the condition in equation~(\ref{eq:wgm-DtN}). 
Therefore, it is natural to infer that the nonzero $\omega''$ of the guided-mode resonances, or perturbed waveguide modes, may arise from their interaction with the FP modes. 
As shown in figure~\ref{fig:free-wg-PhC}(c), the two bands, $\mathrm{GR}_0^{-1}$ and FP$_2$, can intersect at some point, as indicated by the green dashed circle. According to the perturbation theory, the mutual interaction is the strongest at the degeneracy point and decreases with increasing distance from it. This also suggests that the nonzero $\omega''$, although originating from the same perturbation, can vary significantly along the guided-mode resonance band, as detailed below.

Let us rewrite the bulk Bloch states $|\psi_n\rangle$ up to the first order given in equation (\ref{eq:eigenstatepertur-1}) as follows:
\begin{equation}
    |\psi_n\rangle = |n\rangle + u_{n,n_1} |n_1\rangle \delta + u_{n,n_2} |n_2\rangle \delta + \cdots, 
    \label{eq:Blochfunc}
\end{equation}
where the terms on the right-hand side are arranged in the descending order such that $|u_{n,n_1}|\geq |u_{n,n_2}| \geq \cdots$. The simplest nontrivial case considers only two bulk Bloch waves, $\psi_0$ and $\psi_{-1}$ for $q \in (0,\pi/a)$. To determine the weight of these two Bloch waves, it is also necessary to consider two corresponding Fourier components in equation~(\ref{eq:Blochfunc}) as a boundary condition should be imposed for each Fourier component, that is, the number of Fourier components should equal the number of bulk Bloch waves. When polarisation effects are included, the number of background plane waves should account for both the polarisation degrees of freedom and Fourier components, as discussed in section~\ref{sec:GRs-TE&TM}. 

Therefore the bulk Bloch wave functions in equation~(\ref{eq:Blochfunc}) become
\begin{equation*}
\begin{array}{ll}
    |\psi_0\rangle &= |0\rangle + u_{0,{-1}} |{-1}\rangle \delta + \cdots, \\
    |\psi_{{-1}}\rangle &= |{-1}\rangle + u_{{-1},0} |0\rangle \delta + \cdots.
\end{array}
\end{equation*}
The other Fourier components are neglected because the wave functions $\psi_0$ and $\psi_{{-1}}$ are primarily influenced by the $|0\rangle$ and $|$-1$\rangle$ components, and the magnitudes of the perturbation coefficients $u_{nm}$ in equation~(\ref{eq:eigenstatepertur}) decrease as the energy level spacing increases, making the nearest level dominant.  
In equation~(\ref{eq:change-of-basis}), we change the basis from the bulk Bloch-wave functions $\mathbf{c}$ to their Fourier components $\tilde{\mathbf{u}}$. In the following, we use $\mathbf{c}$ in the calculations while discussing the results for the $\tilde{\mathbf{u}}$ representation, and further apply the substitution $(c_0,c_{-1}) \rightarrow \big( c_0 \sin(k_{z0}h/2),c_{-1}\sin(k_{z,-1}h/2) \big)$ for simplicity. The boundary conditions at $z= \pm h/2$ yield
\begin{equation}
    \left[
    \begin{array}{ll}
    f_{00} & u_{{-1},0} f_{{-1},0} \delta \\
    u_{0,{-1}} f_{0,{-1}} \delta & f_{{-1},{-1}}
    \end{array}
    \right]
        \left[
        \begin{array}{ll}
            c_0 \\ c_{{-1}}
        \end{array}
        \right]
        = 0, 
    \label{eq:GRdisp-2modes}
\end{equation}
where
\begin{equation}
    f_{mn} = \frac{1}{Z_{\mathrm{PhC},m}} - \frac{1}{Z_{\mathrm{b},n}},
    \label{eq:f_mn}
\end{equation}
and 
\begin{equation*}
    \frac{1}{Z_{m}} = \frac{H_{m,||}}{E_{m,||}} = \frac{\partial_z\psi_m}{\psi_m},
\end{equation*}
represents the surface admittance or DtN operator acting on the $m$-th eigenstate in either the background medium or bulk PhC.

We note that the vanishing of a diagonal term, either $f_{00} = 0$ or $f_{{-1},{-1}} = 0$, corresponds to the FP-mode or waveguide-mode conditions given in equation~(\ref{eq:wgm-DtN}), respectively. 
The FP mode has a real wavevector $q$ but a complex $\omega$, whereas the waveguide mode has a real wavevector $q-G$ but a purely imaginary $k_z^{\mathrm{b}}$. Therefore, the FP and guided-mode resonances are primarily governed by the Bloch waves $\psi_0$ and $\psi_{{-1}}$, respectively \cite{hu2022global}. 
The correction to the dispersion relation, affecting both the real and imaginary parts, arises from the interaction between different Bloch modes, represented by the off-diagonal terms $u_{0,{-1}}$ and $u_{{-1},0}$. 

For an unperturbed waveguide mode characterised by $(q-G,\omega_0)$, let us consider the first-order correction $\omega=\omega_0+\delta \omega$ due to the perturbation for the specified $q-G$. Similar to the convention of the minus sign in $\omega = \omega' - i \omega''$ adopted earlier, $\delta\omega = \delta\omega' - i \delta\omega''$ is also assumed for consistency. Because $f_{{-1},{-1}}(q-G,\omega_0) = 0$, the guided-mode resonance condition in equation~(\ref{eq:GRdisp-2modes}) simplifies to:
\begin{equation}
    \mathrm{det} 
    \left[
    \begin{array}{ll}
    f_{00} & u_{{-1},0} f_{{-1},0} \delta \\
    u_{0,{-1}} f_{0,{-1}} \delta & \delta\omega \partial_{\omega}f_{{-1},{-1}}
    \end{array}
    \right]
    =0.  
    \label{eq:GRdisp-perturb}
\end{equation}

\noindent Thus, the leading-order correction of the dispersion relation is given by
\begin{equation}
    \delta\omega = - C \delta^2, 
    \label{eq:dw-2band-1}
\end{equation}
which is of the order $\mathrm{O}(\delta^2)$. The proportionality constant
\begin{equation}
    C(q) = \frac{f_{0,{-1}} f_{{-1},0}}{ f_{00} \, \partial_{\omega}f_{{-1},{-1}} } |u_{0,{-1}}|^2,
    \label{eq:C_q}
\end{equation}
is analogous to a structure factor, where the relation $u_{{-1},0} = -u^*_{0,{-1}}$ is used.

\begin{figure}[h]
    \centering    
    \includegraphics[width=0.8\textwidth]{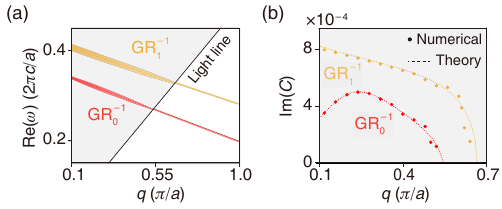}
    \caption{Complex band structure of a 1D PhC slab. (a) Dispersion relation and imaginary part $\omega''$ (spectral width) of guided-mode resonances GR$_0^{-1}$ and GR$_1^{-1}$ at perturbation strength $\delta = 0.1$, with $\omega''$ scaled by a factor of 500 for visibility. (b) Numerical (dots) and theoretical (dashed) results for $C(q)$, defined in equation~(\ref{eq:C_q}) using $\omega'' = C(q)\,\delta^2$.
    }
    \label{fig:Im part of w} 
\end{figure}

The physical significance of equation~(\ref{eq:dw-2band-1}) is as follows. 
First, the correction to the real part $\omega'$ is of order $\mathrm{O}(\delta^2)$, indicating that the interaction between the Bloch waves induces only a small modification to the dispersion relation. 
Figure~\ref{fig:Im part of w}(a) presents the dispersion of the $\mathrm{GR}_0^{-1}$ and $\mathrm{GR}_1^{-1}$ bands for the structure with $h = a$ and perturbation strength $\delta = 0.1$. The spectral width, representing the imaginary part $\omega''$, is also plotted, scaled by a factor of 500 for better visualisation. Second, because $\delta\omega'' = \mathrm{O}(\delta^2)$ rather than $\mathrm{O}(\delta)$, the imaginary part remains non-negative, ensuring the physical stability of the mode, with the proportionality constant $C(q)$ playing a key role.
As shown in figure~\ref{fig:Im part of w}(b), the numerically computed values of $C$ for the two guided-mode resonances agree well with the theoretical predictions derived above from first principles. A more detailed discussion will be presented in the next section. 
Moreover, the $\delta^2$ dependence is reminiscent of quasi-BIC behaviour \cite{liu2019high, koshelev2018asymmetric}, where $\delta$ is an asymmetry parameter such that $\delta = 0$ leads to the formation of a symmetry-protected BIC. 
However, in our case, $\delta$ quantifies the breaking of continuous translational symmetry down to a discrete one. For a small $\delta$, the guided-mode resonances exhibit extremely high-\textit{Q} factors, analogous to those of quasi-BICs \cite{wu2019giant,sun2023infinite}. 
When $\delta = 0$, the continuous symmetry is restored, and all high-\textit{Q} modes revert to the conventional waveguide modes. Finally, because $\omega'' \geq 0$, it follows that $\mathrm{Im}(k_z^{\mathrm{b}}) \leq 0$. With the factor $e^{i k_z^{\mathrm{b}} z}$ in the background medium, the resonant mode solutions grow exponentially in space while decaying exponentially in time. Therefore, they do not represent physical modes but rather correspond to the poles of the $S$-matrix.

\subsection{`Accidental' BIC and its dual FP mode: impedance eigenvalue degeneracy}
\label{sec: Accidental BIC and its dual FP mode}

The theory derived above can be extended to any band of the guided-mode resonances by simply replacing the index $-1$ in equation~(\ref{eq:GRdisp-perturb}) with $m$, the band-folding index of GR$_n^m$. 
The index $n$ is determined using the unperturbed waveguide mode from which the resonance originates. The interaction between the bulk Bloch waves $\psi_0$ and $\psi_m$ leads to the coupling between the FP$_l$ and GR$_n^m$ modes, where $l$ and $n$ denote the number of nodes along the $z$-direction, and $l \equiv n \ (\mathrm{mod}\ 2)$, indicating that these two modes share the same symmetry under the $\sigma_h$ mirror operation. Some special cases may require separate treatment when the two-band model is insufficient. For example, if the first Fourier component of the perturbation $\varepsilon(x)$ vanishes, i.e. $\langle 0 | \varepsilon | {-1} \rangle = 0$, the coefficient $u_{{-1},0}$ in the wavefunction expansion of equation~(\ref{eq:Blochfunc}) also vanishes~\cite{lee2021metasurfaces}. 
This results in the decoupling of the FP and GR modes, such as FP$_2$ and GR$_0^{-1}$. In such cases, the estimation of $\omega''$ requires the inclusion of second- or higher-order corrections in the perturbation theory. Another case involves the effect of a substrate. The breaking of the $x$-$y$ mirror symmetry can induce coupling between modes of opposite parity, such as FP$_1$ and GR$_0^{-1}$.~\cite{lee2020bound}. 
Nevertheless, these special cases can also be addressed within our framework. For example, in supplementary section 1, the case with a substrate is treated. Due to the $\sigma_h$ symmetry breaking, the upward and downward propagating Bloch waves become linearly independent, leading to a doubled dimension of Hilbert space, as well as a doubled number of scattering channels. However, the underlying principle remains unchanged: the minimal Hilbert space dimension is still determined solely by the number of propagating bulk Bloch waves, which includes both the upward and downward propagating Bloch waves.

\begin{figure}[h]
    \centering    
    \includegraphics[width=0.8\textwidth]{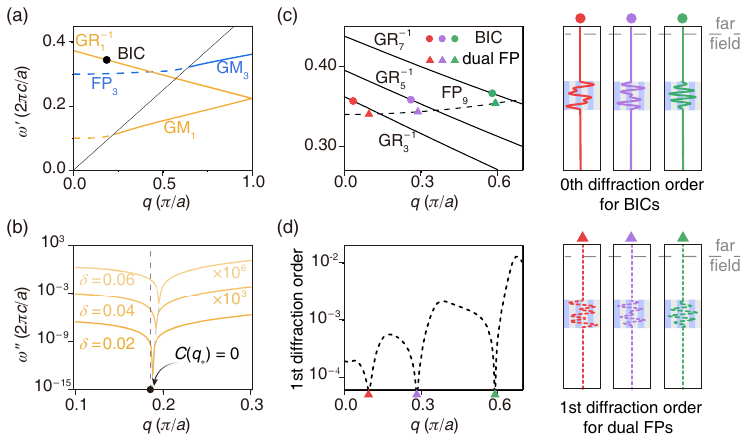}
    \caption{Accidental BICs and their dual FP modes. (a) An accidental BIC (black dot) arises from the coupling between GR$_1^{-1}$ and FP$_3$ bands (with $\delta=0.02$, $h=1.67a$). (b) Imaginary part of the frequency for GR$_1^{-1}$ at $\delta=0.02$, 0.04, 0.06: the BIC converges to a fixed point as $\delta\to0$. (c) Accidental BICs and dual FP modes arising from impedance eigenvalue degeneracy (for $\delta=0.01$, $h=4.42a$). (d) First order diffraction amplitude on FP$_9$: dual FP modes have zero first order diffraction, whereas BICs have zero zeroth order diffraction. Right panels: electric-field profiles of zeroth and first diffraction components for BICs (top) and dual FPs (bottom).    }
    \label{fig:accidental BIC and dual FP} 
\end{figure}

Let us take the GR$_1^{-1}$ band in figure~\ref{fig:accidental BIC and dual FP}(a) as an example. It intersects with the FP$_3$ mode (blue dashed line), near which the interaction between FP$_3$ and GR$_1^{-1}$ is significantly enhanced, allowing the formation of a BIC~\cite{hu2022global}. 
By definition, a BIC should have a real-valued eigenfrequency, and the imaginary part of the eigenfrequency obtained from equation~(\ref{eq:dw-2band-1}) should vanish, as follows:
\begin{equation}
    \delta\omega''=0,
    \label{eq:BIC-Im-w-1st}
\end{equation}
\noindent implying that $u_{0,{-1}}=0$ or $C$ is purely real. The case $u_{0,{-1}}=0$ has been discussed above; it corresponds to the condition $\langle {-1}|\varepsilon(x)|0 \rangle = 0$, i.e. the perturbation does not couple the two unperturbed states. In this case, $\delta\omega'' = 0$ is trivial.

Another possibility is that $C \in \mathbb{R}$, which includes several subcases. Noting that $f_{0,{-1}} \in \mathbb{R}$, a straightforward condition is $f_{0,{-1}} = 0$. Together with the waveguide condition for the unperturbed state $|{-1}\rangle$, i.e. $f_{{-1},{-1}} = 0$, this leads to the condition
\[
k_{z0} \cot\left( k_{z0}h/2 \right) = k_{z,{-1}} \cot\left( k_{z,{-1}}h/2 \right).
\]
In a more general form, this condition can be written as
\begin{equation}
\left.\frac{\partial_z \Psi}{\Psi}\right|_{\mathrm{mode\ 0}} = \left.\frac{\partial_z \Psi}{\Psi}\right|_{\mathrm{mode\ {-1}}},
\label{eq:wgm-DtNdegeneracy}
\end{equation}
which means that the DtN operator, or surface impedance matrix, has degenerate eigenvalues for the two bulk Bloch waves indexed by 0 and $-1$. These two Bloch waves share the same frequency $\omega$ but differ in their wavevectors $q$ and $q-G$.
Because the surface impedance depends on the slab thickness $h$, equation~(\ref{eq:wgm-DtNdegeneracy}) can be used to determine $\omega$ for fixed $q$ and $h$, or to solve for any one of them given the other two. This degeneracy can occur at arbitrary points in the $(q, \omega)$ space, which is why such BICs are referred to as `accidental'. 
When a twofold degeneracy exists in the eigenvalues of the DtN matrix, the unitary transformation $\mathcal{U}$ in equation~(\ref{eq:DtN^PhC vs DtN^b}) leaves the diagonal form of the DtN matrix unchanged in this 2D subspace and preserves the original eigenfrequency $\omega$, thereby forming a BIC.

Another possibility for $C \in \mathbb{R}$ is that the ratio $f_{{-1},0}/f_{00}$ is real, even though both $f_{00}$ and $f_{{-1},0}$ are complex. This corresponds to the condition $\arg(f_{{-1},0}/f_{00}) = 0$, which leads to the same result as equation~(\ref{eq:wgm-DtNdegeneracy}).

One important question to be addressed is whether the sign of $\delta\omega''$ changes as it passes through zero. If it does, the resulting solution would become unphysical, exhibiting an exponential growth in time. The positive semi-definiteness of $\omega''$ can also be inferred from the definition of the quality factor
\begin{equation}
    Q = \frac{\omega'}{2\omega''} = \omega' \cdot \frac{\mathrm{Energy\ stored}}{\mathrm{Power\ radiated}} \geq 0. \label{eq:Qdef}
\end{equation}

\noindent Let us assume that equation~(\ref{eq:wgm-DtNdegeneracy}) is satisfied at $q = q_*$ and $\omega = \omega_*$. As previously discussed, both conditions $f_{0,-1} = 0$ and $\arg(f_{-1,0}/f_{00}) = 0$ lead to the same result, namely, equation~(\ref{eq:wgm-DtNdegeneracy}). 
As $q$ varies from $q_*^{-}$ to $q_*^{+}$, both $\arg(f_{-1,0}/f_{00})$ and $f_{0,-1}$ undergo a sign reversal. The combined effect of these two conditions ensures $\delta\omega'' \geq 0$ in the vicinity of $q = q_*$. This is a subtle yet crucial aspect of the nontrivial BIC condition.

To form this type of BIC, the behaviour of the DtN operator acting on the eigenstates of the PhC and background medium is critical. For a uniform dielectric slab, the DtN matrix evaluated at $\epsilon = \bar{\epsilon}$ commutes with that in the background medium, i.e. DtN$|_{\epsilon_\mathrm{b}}$, allowing the FP and waveguide modes to be determined straightforwardly. However, for a PhC slab, the DtN operator with nonzero perturbation ($\delta \neq 0$) no longer commutes with that of the background, thereby enabling coupling between different bulk Bloch waves. When two eigenvalues of DtN$|_{\delta \neq 0}$ become degenerate, a BIC can emerge, where the second-order correction to $\delta\omega$ vanishes. Therefore, this type of BIC can be interpreted as a fixed point of perturbation, that is,
\begin{equation*}
    \frac{\partial (\delta\omega)}{\partial \delta} = \frac{\partial^2 (\delta\omega)}{\partial \delta^2} = 0.
\end{equation*}

The fixed-point nature of the BIC in the limit $\delta \rightarrow 0$ can also be seen from equation~(\ref{eq:wgm-DtNdegeneracy}), the solution of which is independent of the perturbation strength $\delta$. 
Figure~\ref{fig:accidental BIC and dual FP}(b) illustrates the imaginary part $\omega''$ of guided-mode resonances.
As $\delta \rightarrow 0$, the zero of $\omega''$ asymptotically approaches a fixed point. This type of BIC is referred to as an `accidental' BIC \cite{hsu13,bulgakov2019high} for two reasons: (1) its Bloch wavevector $q$ can be tuned using geometric parameters such as the slab thickness $h$, and (2) it appears to stem from a single-mode resonance when the FP modes are considered merely as background in the transmission spectrum.
Besides accidental BICs, one can clearly see that the imaginary part remains small across the entire band due to the small perturbation strength $\delta$ employed. As $\delta$ increases, the $Q$ factor of the band decreases significantly. See supplementary section 2 for details.

These interpretations are also supported by equation~(\ref{eq:wgm-DtNdegeneracy}). 
For the first reason, the solution $(q, \omega)$ depends on $h$ and is not constrained to specific symmetry points in the $q$-space. For the second, the solution $(q, \omega)$ may lie far from the crossing point between the FP and folded waveguide-mode bands. More importantly, the eigenvalue degeneracy of the impedance matrix reflects an intrinsic property of the bulk Bloch states in periodic media. Therefore, the reduction in the translational symmetry—from continuous to discrete—fundamentally enables the formation of BICs.

The degeneracy of the impedance matrix can lead to intriguing physical effects. By examining the corresponding wave function obtained from equation~(\ref{eq:GRdisp-2modes}),
\begin{equation*}
    \textbf{c} \approx \left( u_{{-1},0} \delta / \sin(k_{z0}h/2),\ -1/\sin(k_{z,{-1}}h/2) \right)^\mathrm{T},
\end{equation*}
the resulting expressions for $\Psi$ and $\partial_z\Psi$ are
\begin{equation*}
\begin{array}{ll}
     &  \Psi = -\frac{\mathrm{sin}(k_{z,{-1}}z)}{\mathrm{sin}(k_{z,{-1}}h/2)} |{-1}\rangle + \left( \frac{\mathrm{sin}(k_{z,0}\,z)}{\mathrm{sin}(k_{z0}h/2)} - \frac{\mathrm{sin}(k_{z,{-1}}\,z)}{\mathrm{sin}(k_{z,{-1}}h/2)} \right) u_{{-1},0} \delta |0\rangle, \\
     & \partial_z \Psi = -\frac{ k_{z,{-1}} \mathrm{cos}(k_{z,{-1}}z)}{\mathrm{sin}(k_{z,{-1}}h/2)} |{-1}\rangle + \left( \frac{ k_{z,0} \mathrm{cos}(k_{z,0}\,z)}{\mathrm{sin}(k_{z0}h/2)} - \frac{ k_{z,{-1}} \mathrm{cos}(k_{z,{-1}}\,z)}{\mathrm{sin}(k_{z,{-1}}h/2)} \right) u_{{-1},0} \delta |0\rangle.
\end{array}    
\end{equation*}
It is evident that the degeneracy of the impedance matrix makes it possible for the $|0\rangle$ components on the right-hand side of the above equations to vanish at $z=h/2$. The same result can be obtained for the interface at $z=-h/2$ owing to the $\sigma_h$ mirror symmetry.
This corresponds to the formation of a BIC, as illustrated in figure~\ref{fig:accidental BIC and dual FP}(c). 
Furthermore, an additional solution is given by 
\begin{equation*}
   {\mathbf{c}} \approx \big(1/\sin({k_{z,0}}h/2),\ - u_{0,{-1}}\delta /\sin({k_{z,{-1}}}h/2) \big)^{\rm{T}}.
\end{equation*} 
\nonumber In this case, the $|{-1}\rangle$ components of both $\Psi$ and $\partial_z \Psi$ vanish at $z = \pm h/2$, also owing to the impedance eigenvalue degeneracy.

In contrast to the BIC, this degeneracy occurs on the FP band rather than the guided-mode resonance band, giving rise to an FP mode that is dual to the `accidental' BIC. However, unlike the BIC, which possesses a real-valued eigenfrequency, the dual FP mode exhibits a complex frequency, making it more difficult to strictly satisfy the impedance degeneracy condition. This limitation can be overcome by fine-tuning parameters such as the slab thickness $h$.
As shown in figure~\ref{fig:accidental BIC and dual FP}(c) and (d), both the BICs and their dual FP modes (indicated by triangles) are presented for FP$_9$ and the corresponding three guided-resonance bands. The disappearance of the Fourier component $|{-1}\rangle$ outside the slab can be observed in the right panel of figure~\ref{fig:accidental BIC and dual FP}(d). 
In comparison with other FP modes, these dual FP modes exhibit significantly stronger spatial confinement, as their near fields are contributed by Fourier components higher than $|{-1}\rangle$.

Furthermore, the proportionality constant $C$ in equation~(\ref{eq:C_q}) can be expanded near $q = q_*$ as
\begin{equation}
    C(q) = \tilde{C} \cdot (q - q_*)^2,
    \label{eq:C-tilde}
\end{equation}
given that $C(q_*) = 0$. Because we focus on the imaginary part of $C(q)$, the replacement $\tilde{C} \rightarrow \mathrm{Im}({\tilde{C}})$ is assumed below. This has the same functional form as that in reference~\cite{yuan2017strong}. 
However, in contrast to the previous approach, our analysis is grounded in a first-principles $S$-matrix formalism. Within this framework, $\tilde{C}$ can be rigorously evaluated for any arbitrary $q_*$. It is given by
\begin{equation*}
\tilde{C} = -\left[\partial_q \left( \frac{1}{Z_{\rm{PhC},-1}} - \frac{1}{Z_{\rm{PhC},0}} \right) \right]^2 \cdot |u_{0,-1}|^2 \Big/ \left( i Z_{\rm{b},0} |f_{00}|^2 \, \partial_\omega f_{-1,-1} \right) \Big|_{q=q_*},
\end{equation*}
which depends only on the structural parameters such as the period, slab thickness, and effective permittivity.
According to the proportionality coefficient in equation (27), the $Q$ factor near the BIC follows the scaling law $Q \propto 1/(q - q_*)^2$. This inverse-square dependence is validated by the numerical simulations presented in supplementary section 2.

\subsection{Friedrich--Wintgen and symmetry-protected BICs: interaction of guided-mode resonances}\label{sec:FW-BIC}

The two-band model described above effectively explains the origin of $\omega''$ and resolves most problems concerning guided-mode resonances. However, there exists another type of crossing point between distinct guided-mode resonance branches, such as the GR$^{1}_0$ and GR$^{-1}_2$ bands in figure~\ref{fig:free-wg-PhC}(c), highlighted by the green circle, near which the applicability of the two-band model must be reconsidered. 
Notably, these two resonant modes originate from different waveguide modes, as reflected by their indices. For example, GR$^{1}_0$ arises from a waveguide mode without any node along the \textit{z}-direction in the slab and is folded twice into the first BZ, whereas GR$^{-1}_2$ stems from a waveguide mode with two nodes and is folded once. 

A three-band model is required to account for the interaction of both guided-mode resonances with the FP modes, particularly when they couple to the same FP mode. In other words, for a PhC slab that is infinite in the $x$-$y$ plane, all FP and guided-mode resonances form 2D band surfaces in the $(\textbf{k}_{||}, \omega)$ space. Intersections between these surfaces, i.e. lines where two surfaces cross, are inevitable, as discussed in detail in section~\ref{sec: Accidental BIC and its dual FP mode}. 
Moreover, the intersection of three such surfaces is also a generic situation, resulting in an intersection point rather than a line, which corresponds to the scenario investigated here. Intersections involving four or more surfaces are highly unlikely, unless additional degeneracies arise from polarisation or spatial symmetries in the 2D structure, as discussed later.

In the three-band model, the wave function within the slab, described by equation~(\ref{eq:wavefunc-slab}), consists of three bulk Bloch waves, as follows:
\begin{equation}
   \Psi = a_0 \cos(k_{z0}z)\psi_{0} + a_{-1} \cos(k_{z,-1}z)\psi_{-1} + a_{1} \cos(k_{z1}z)\psi_{1}. 
   \label{eq:3-Bloch-waves}
\end{equation}
For convenience, we restrict $q \in (0,\pi/a)$. The Bloch-wave functions are expanded in a Fourier basis as
\begin{equation}
\begin{array}{lll}
    |\psi_0\rangle &= |0\rangle + u_{0,-1}|{-1}\rangle \delta + u_{0,1}|1\rangle \delta + \cdots, \\
    |\psi_{-1}\rangle &= |{-1}\rangle + u_{-1,0}|0\rangle \delta + u_{-1,1}|1\rangle \delta + \cdots, \\
    |\psi_1\rangle &= |1\rangle + u_{1,0}|0\rangle \delta + u_{1,-1}|{-1}\rangle \delta + \cdots,
\end{array}
\end{equation}
where we assume $|u_{0,-1}| \geq |u_{0,1}|$, $|u_{-1,0}| \geq |u_{-1,1}|$, and $|u_{1,0}| \geq |u_{1,-1}|$, implying the terms are arranged in the descending order of significance. All off-diagonal terms $u_{mn} = \mathrm{O}(1)$ for $m \ne n$, with the perturbation strength $\delta$, are factored out. Terms such as $u_{0,1}$, $u_{-1,1}$, and $u_{1,-1}$ represent the next-nearest-neighbour contributions.

Recalling the auxiliary function $f_{mn} = 1/Z_{\mathrm{PhC},m} - 1/Z_{\mathrm{b},n}$ defined in equation~(\ref{eq:f_mn}), we note that the conditions $f_{-1,-1} = f_{1,1} = 0$ are satisfied at the crossing point $(q_c, \omega_c)$ of the two waveguide modes. Near $q_c$, with $\omega = \omega_c + \delta \omega$, matching boundary conditions at $z = \pm h/2$ yields the following relation:
\begin{equation}
    \mathrm{det} \left[
    \begin{array}{lll}
        f_{00} & u_{-1,0} f_{-1,0} \delta & u_{1,0} f_{1,0} \delta \\
        u_{0,-1} f_{0,-1} \delta & \delta\omega \partial_{\omega}f_{-1,-1} & u_{1,-1} f_{1,-1} \delta \\
        u_{0,1} f_{0,1} \delta & u_{-1,1} f_{-1,1} \delta & \delta\omega \partial_{\omega}f_{1,1}
    \end{array} \right]
    = 0.
    \label{eq:GRdisp-3band-cross}
\end{equation}

\begin{figure}[h]
    \centering    
    \includegraphics[width=0.8\textwidth]{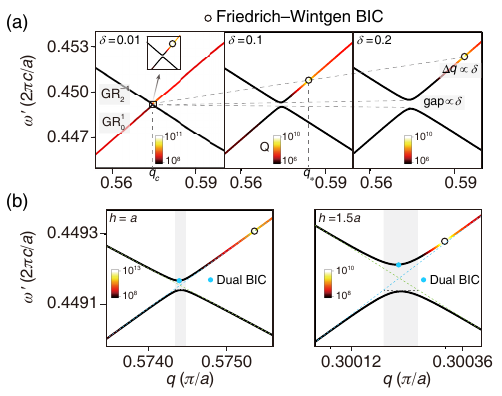}
    \caption{Friedrich--Wintgen BICs and their duals. (a) Dispersion of GR$_0^1$ band shows a Friedrich--Wintgen BIC arising from its coupling with GR$_2^{-1}$ at $\delta=0.01$, 0.1, 0.2 and $h=a$. The bandgap and BIC position ($\Delta q = q_* - q_c$, where $q_c$ is the band crossing point) each scale linearly with $\delta$. (b) The dual BIC emerges on GR$_2^{-1}$ near the crossing point, but is disrupted by mode hybridization. 
    }
    \label{fig:FW-BIC} 
\end{figure}

\noindent This condition at the point $q_c$ takes the form of a quadratic equation
\begin{equation}
    a\, \delta\omega^2 + b\, \delta^2 \delta\omega + c\, \delta^2 = 0, 
    \label{eq:GRdisp-3band-cross1}
\end{equation}
where
\begin{equation*}
\begin{array}{lll}
    a &= f_{00} \left( \partial_{\omega}f_{-1,-1} \right) \left( \partial_{\omega}f_{1,1} \right) = \mathrm{O}(1), \\
    b &= - u_{0,1} u_{1,0} f_{0,1} f_{1,0} \left( \partial_{\omega}f_{-1,-1} \right) - u_{0,{-1}} u_{{-1},0} f_{0,{-1}} f_{{-1},0} \left( \partial_{\omega}f_{1,1} \right) = \mathrm{O}(1), \\
    c &= f_{00} u_{-1,0} u_{1,-1} f_{-1,0} f_{1,-1} + \cdots = \mathrm{O}(1).
\end{array}
\end{equation*}
Then, the solution is
\begin{equation}
    \delta\omega = \frac{-b \delta^2 \pm \sqrt{b^2 \delta^2 - 4ac} \,\delta }{2a}. 
    \label{eq:GRdisp-3band-cross2}
\end{equation}
The first term, $-(b/2a)\delta^2$, corresponds to a uniform shift of the crossing point, whereas the second term, proportional to $\delta$, indicates an avoided level crossing.
As shown in figure~\ref{fig:FW-BIC}(a), a bandgap of the order $\mathrm{O}(\delta)$ opens between GR$_0^{1}$ and GR$_2^{-1}$, as marked by the dashed lines. A bandgap does not open if $b^2\delta^2 - 4ac \leq 0$, in which case the square root is zero or purely imaginary, contributing no gap to the real part of the frequency. Nevertheless, satisfying such a constraint is nontrivial, as the matrix elements in the equation~(\ref{eq:GRdisp-3band-cross}) are generally complex. 
This explains the frequent observation of the level repulsion phenomenon. 

It is known that near the avoided crossing point of two guided-mode resonances, a peculiar point may arise where a Friedrich--Wintgen BIC can be formed \cite{fried85, kang2021merging}. 
To determine the position of such a BIC, we slightly move away from the crossing point and expand equation~(\ref{eq:GRdisp-3band-cross}) in the neighbourhood of $(q_c, \omega_c)$. 
Let $(q, \omega) = (q_c + \tilde{q} \delta, \omega_c + \tilde{\omega} \delta + \delta\omega)$, and assume that the BIC lies on the band indexed by superscript 1 (an analogous analysis applies to the $-1$ band). Then, we have
\begin{equation*}
    f_{11}(q_c, \omega_c) = 0, \qquad
    f_{11}(q_c + \tilde{q} \delta, \omega_c + \tilde{\omega} \delta) = 0.
\end{equation*}

\noindent
These two expressions represent the linear approximation of the dispersion relation for the unperturbed band 1. $\delta\omega$ denotes the frequency deviation from this unperturbed waveguide mode. Substituting into equation~(\ref{eq:GRdisp-3band-cross}) yields:
\begin{equation}
    \mathrm{det} \left[
    \begin{array}{ccc}
        f_{00} & u_{{-1},0} f_{{-1},0} \delta & u_{1,0} f_{1,0} \delta \\
        u_{0,{-1}} f_{0,{-1}} \delta & f'_{{-1},{-1}} \delta & u_{1,{-1}}f_{1,{-1}} \delta \\
        u_{0,1} f_{0,1} \delta & u_{{-1},1}f_{{-1},1} \delta & (\partial_{\omega}f_{11})\delta\omega 
    \end{array} \right]
    =0, 
    \label{eq:FW-BIC}
\end{equation}
where $f'_{{-1},{-1}} \equiv \big( f_{{-1,-1}}(q,\omega) - f_{-1,-1}(q_c,\omega_c) \big)/\delta = \tilde{q} \, \partial_q f_{-1,-1} = \mathrm{O}(1)$. At the crossing point $(q_c, \omega_c)$, the waveguide conditions $f_{{-1},{-1}} = f_{1,1} = 0$ hold. We define the following auxiliary quantities:
\begin{equation*}
\begin{array}{lll}
    a' &= (f_{00} f'_{-1,-1} \, \partial_{\omega} f_{11}) \delta + (|u_{-1,0}|^2 f_{-1,0} f_{0,-1} \, \partial_{\omega} f_{11}) \delta^2 \equiv a'_1 \delta + a'_2 \delta^2, \\
    b' &= |u_{-1,1}|^2 f_{00} f_{-1,1} f_{1,-1}, \\
    c' &= u_{0,1} u_{-1,0} u_{1,-1} (f_{0,1} f_{-1,0} f_{1,-1} - f_{0,-1} f_{-1,1} f_{1,0}) + |u_{0,1}|^2 f'_{-1,-1} f_{1,0} f_{0,1}.
\end{array}
\end{equation*}

\noindent
Assuming the periodic perturbation $\varepsilon(x)$ is even-symmetric and the coordinate origin is chosen on the symmetry axis, the mode-coupling coefficients $u_{mn}$ are real-valued, i.e., $u_{mn}\in \mathbb{R}$. Combined with the anti-Hermiticity of the matrix $\mathbf{U}$, this implies $u_{mn} = -u_{nm}$. Solving equation~(\ref{eq:FW-BIC}) and expanding $\delta\omega$ in a Taylor series around $\delta = 0$ gives
\begin{equation}
    \delta\omega = -\frac{b'}{a'_1} \delta + \frac{a'_2 b' - a'_1 c'}{a_1'^2} \delta^2 + \mathrm{O}(\delta^3).
    \label{eq:FWBIC-2}
\end{equation}

\noindent The leading-order term $-b'/a'_1 \cdot \delta$ is real-valued, as all $f$-functions are real near the crossing point, implying that the first-order correction to $\delta\omega$ is real. The second-order term involves complex quantities and corresponds to the proportionality constant $C(q)$ defined in equation~(\ref{eq:C_q}). 
Because $f'_{-1,-1} \propto \tilde{q}$ and both $a'_1$ and $c'$ depend on $f'_{-1,-1}$, the BIC condition requires that the second-order correction term in equation~(\ref{eq:FWBIC-2}) is real, as follows:
\begin{equation*}
    \arg\left( a'_2 b' - a'_1 c' \right) = \arg(f_{00}^2),
\end{equation*}
where $a'_1, c' \propto f'_{-1,-1}$. This is the Friedrich--Wintgen BIC condition, which reduces to a quadratic equation in $f'_{-1,-1}$. Remarkably, this equation has a single real root of multiplicity two (see supplementary section 3), implying that the BIC condition is uniquely satisfied. The explicit solution is
\begin{equation}
    \tilde{q} = \frac{u_{0,-1} u_{-1,1} \left[ f_{0,-1} f_{-1,1} \, \mathrm{Im}(f_{1,0}/f_{00}) - f_{0,1} f_{1,-1} \, \mathrm{Im}(f_{-1,0}/f_{00}) \right]}{2 u_{0,1} f_{0,1} (\partial_q f_{-1,-1}) \, \mathrm{Im}(f_{1,0}/f_{00})},
    \label{eq:FW-BIC-cond-SMat}
\end{equation}
where we use $f'_{-1,-1} = \tilde{q} \, \partial_q f_{-1,-1}$. If the mirror symmetry is preserved while the origin is shifted, $u_{mn}$ acquires an additional phase factor, but the Friedrich--Wintgen BIC condition can still be satisfied (see supplementary section 3). Therefore, a Friedrich--Wintgen BIC is formed for a specific value of $\tilde{q}$ that satisfies this condition, indicating that the deviation of the BIC from the crossing point is proportional to $\delta$ with a proportionality constant $\tilde{q}$. These BICs are marked by circles on the GR$_0^1$ band in figure~\ref{fig:FW-BIC}(a), where the proportional dependence on $\delta$ is evident and shows excellent agreement with the numerical simulations. 

When the relative permittivity lacks even symmetry, the coupling coefficients $u_{mn}$ become complex. A true Friedrich--Wintgen BIC is not guaranteed and typically does not exist (see supplementary section 3). This is analogous to that of accidental BICs or symmetry-protected BICs under symmetry breaking~\cite{carletti2018giant, liu2019high, liu2019circularly, chen2023observation, kuhner2023unlocking, sun2023infinite}. This theoretical framework, derived from a three-band $S$-matrix model under symmetry preserving, demonstrates that Friedrich--Wintgen BICs generally exist near the avoided crossing of guided-mode resonances, consistent with the interpretations based on the effective two-level Hamiltonians in previous studies \cite{fried85, sadr21, hu2022global, hsu2016bic, volya2003nonHermi}. 

We briefly introduce the following effective non-Hermitian Hamiltonian for two coupled resonant modes:
\begin{equation*}
    H = 
    \left[
    \begin{array}{cc}
        \omega_1 & \kappa \\
        \kappa & \omega_2 \\
    \end{array}
    \right]
    -i
    \left[
    \begin{array}{cc}
    \gamma_1 & \pm\sqrt{\gamma_1\gamma_2} \\
    \pm\sqrt{\gamma_1\gamma_2} & \gamma_2 \\
    \end{array}
    \right],
\end{equation*}
where $\omega_i$ and $\gamma_i$ are the real and imaginary parts of the $i$-th mode resonant frequency, respectively, and $\kappa$ and $\pm\sqrt{\gamma_1\gamma_2}$ are the near-field and far-field coupling terms. The $\pm$ sign corresponds to even (odd) symmetry with respect to the $x$-$y$ mirror plane. Then, the condition for a Friedrich--Wintgen BIC is
\begin{equation}
    \kappa (\gamma_1 - \gamma_2) = \pm \sqrt{\gamma_1 \gamma_2} (\omega_1 - \omega_2),
    \label{eq:FW-BIC-cond-Heff}
\end{equation}
which yields a single solution under the assumption that $\gamma_i$ are approximately constant near the band-crossing point. The choice of the branch---whether the BIC occurs to the left or right of the crossing---depends on the signs of $\omega_1 - \omega_2$, $\gamma_1 - \gamma_2$, and $\kappa$, and the mode symmetry. If $\kappa \propto \delta$ is assumed, the band gap will be proportional to $\delta$.

However, the three-band model exhibits more intricate physics. In this model, it is not necessary to assume that the BIC lies on the GR$_0^1$ band. A similar analysis can be conducted for the GR$_2^{-1}$ band, and the condition becomes
\begin{equation*}
    \tilde{q} = \frac{u_{0,1} u_{1,-1} [f_{0,-1} f_{-1,1} \mathrm{Im}(f_{10}/f_{00}) - f_{0,1} f_{1,-1} \mathrm{Im}(f_{-1,0}/f_{00}) ]}{2 u_{-1,0} f_{0,-1} (\partial_q f_{11}) \mathrm{Im}(f_{-1,0}/f_{00})}.
\end{equation*}
This condition for the Friedrich--Wintgen BIC on the GR$_2^{-1}$ band reveals a duality with that on the GR$_0^{1}$ band.
As shown in figure~\ref{fig:FW-BIC}(b), the dual solutions on the GR$_2^{-1}$ band are indicated by blue dots. These dual BICs appear very close to the crossing point between the GR$_0^{1}$ and GR$_2^{-1}$ bands. To quantitatively estimate their proximity to the crossing point, the region $\Delta k$ corresponding to the width of the band gap is highlighted in grey in figure~\ref{fig:FW-BIC}(b). 
It is evident that the dual Friedrich--Wintgen BIC lies within this grey region. Unlike the original Friedrich--Wintgen BIC (denoted by a circle), the dual BIC does not exhibit a divergent-$Q$ factor. Although this solution always exists and represents the dual counterpart of a Friedrich--Wintgen BIC, it fails to exhibit the defining characteristics of a true BIC.

This observation raises a question regarding the dual BIC solutions in the two-band and three-band models: why does the dual FP mode always exist in the two-band model, whereas the dual BIC is destroyed in the three-band case? In the two-band model, the degeneracy of the impedance matrix can be satisfied on both the FP and guided-mode resonance bands, even far from their crossing point, allowing the perturbation theory to remain valid. In contrast, in the three-band model, the dual BIC occurs near the crossing point of two guided-mode resonance bands, where the interaction is strong and the perturbation theory based on a single band becomes insufficient. In such cases, both bands must be included in the analysis. Thus, the dual BIC is destroyed by the mixing of the two guided-mode resonances, which aligns with the predictions from an effective non-Hermitian Hamiltonian involving only two coupled energy levels.
If the interaction between the two guided-mode resonances does not directly lead to avoided crossing, the dual solution may exhibit the characteristic of a high $Q$ factor. However, this typically requires precise tuning of structural parameters to close the band gap. Realizing a real dual BIC remains a challenge.

In comparison with this phenomenological effective Hamiltonian model, the three-band model derived from first principles offers a more detailed physical picture. It explicitly shows how the band gap depends on the perturbation strength $\delta$.
For example, as shown in equation~(\ref{eq:GRdisp-3band-cross2}), the interaction between the two guided-mode resonance bands leads to an avoided level crossing, with a band gap proportional to $\delta$. Additionally, both the guided-mode resonance bands undergo a simultaneous shift proportional to $\delta^2$, resulting from their coupling to the FP mode. More importantly, we demonstrate that the Friedrich--Wintgen BIC originates from the crossing of the two guided-mode resonance bands and deviates from the crossing point by an amount proportional to $\delta$, with the proportionality constant $\tilde{q}$ given in equation~(\ref{eq:FW-BIC-cond-SMat}). 
cannot be captured by the effective non-Hermitian Hamiltonian model.

In addition to the Friedrich--Wintgen BICs, symmetry-protected BICs can be interpreted using the above three-band model. These BICs appear at high-symmetry points in the BZ, such as the $\Gamma$ point, and originate from the interaction between degenerate guided-mode resonances. Therefore, they can be treated as a special class of Friedrich--Wintgen BICs. However, in contrast to typical Friedrich--Wintgen BICs, which deviate from the crossing point, symmetry-protected BICs always appear at the high-symmetry point and can be analysed using the perturbation theory for the degenerate case, similar to that discussed in equation~(\ref{eq:GRdisp-3band-cross}).

For simplicity, we consider a twofold degeneracy, where the states $|m\rangle$ and $|{-m}\rangle$ form a degenerate subspace, with $m$ also denoting the band-folding index. In comparison with the non-degenerate case shown in equations~(\ref{eq:eigenvaluepertur}) and (\ref{eq:eigenstatepertur}), the resulting eigenvalues and eigenstates are given by
\begin{equation}
\begin{array}{ll}
\lambda_\pm &= \bar{\epsilon}k_0^2 - m^2 G^2 \pm k_0^2 |\langle -m|\varepsilon(x)|m\rangle| \delta + \mathrm{O}(\delta^2), \\
|\psi_\pm\rangle &= |\psi^{(0)}_\pm\rangle + \Big( u_\mp |\psi^{(0)}_\mp\rangle + \sum\limits_{n \neq \pm} u_{\pm,n} |n\rangle \Big) \delta + \mathrm{O}(\delta^2),
\end{array}
\label{eq:eigen-sys-pertur-degeneracy}
\end{equation}
where $u_{\pm,n} = k_0^2 \langle n|\varepsilon(x)|\psi^{(0)}_\pm \rangle/(n^2 - m^2)G^2$, and $|\psi^{(0)}_\pm \rangle = \frac{1}{\sqrt{2}} (|{-m}\rangle \pm |m\rangle)$. Note that the correction with the eigenvalues $\lambda_\pm$ (i.e. $k_{z,\pm}^2$) is of the order $\delta$, in contrast to that in the non-degenerate case where the correction is of the order $\delta^2$, as shown in equation~(\ref{eq:eigenvaluepertur}). 
The perturbative contributions from other non-degenerate states are treated in the same manner as in equation~(\ref{eq:eigenstatepertur-1}).

For the case of $m=\pm1$, the Bloch wavefunctions can be expanded in the Fourier series as follows:
\begin{equation}
\begin{array}{lll}
|\psi_0\rangle &= |0\rangle + u_{0,-1}|{-1}\rangle \delta + u_{0,1}|1\rangle \delta + \cdots, \\
|\psi_{-}\rangle &= \frac{1}{\sqrt{2}} (|{-1}\rangle - |1\rangle) + u_{-,0} |0\rangle \delta + \cdots, \\
|\psi_{+}\rangle &= \frac{1}{\sqrt{2}} (|{-1}\rangle + |1\rangle ) + u_{+,0}|0\rangle \delta  + \cdots.
\end{array}
\end{equation}

\noindent Recalling the auxiliary functions $f_{mn} = 1/Z_{\mathrm{PhC},m} - 1/Z_{\mathrm{b},n}$ defined in equation~(\ref{eq:f_mn}), we note that the conditions $f_{-1,-1} = f_{1,1} = 0$ are satisfied at the crossing point $(q_c,\omega_c)$. Furthermore, because we focus on the degenerate states at the $\Gamma$ point, where $q_c = 0$, it follows directly that $f_{-1,1} = f_{1,-1} = 0$. Performing a Taylor expansion in $\omega = \omega_c + \delta\omega$ and $\delta$, and applying the boundary conditions at $z = \pm h/2$, we obtain

\begin{equation}
    \mathrm{det}
    \left[
    \begin{array}{ccc}
        f_{00} & u_{-,0}f_{{-1},0} \delta & u_{+,0}f_{1,0} \delta \\
        u_{0,{-1}}f_{0,{-1}} \delta & \frac{1}{\sqrt{2}}(\delta\omega\partial_\omega {+} \delta\partial_\delta) f_{{-1},{-1}} & \frac{1}{\sqrt{2}}(\delta\omega\partial_\omega {+} \delta\partial_\delta) f_{1,{-1}} \\
        u_{0,1}f_{0,1} \delta & -\frac{1}{\sqrt{2}}(\delta\omega\partial_\omega {+} \delta\partial_\delta) f_{{-1},1} & \frac{1}{\sqrt{2}}(\delta\omega\partial_\omega {+}\delta\partial_\delta) f_{1,1} \\
    \end{array}
    \right]
    \label{eq:GR-at-Gamma}
     = 0.
    \end{equation}

\begin{figure}[h]
    \centering    
    \includegraphics[width=0.8\textwidth]{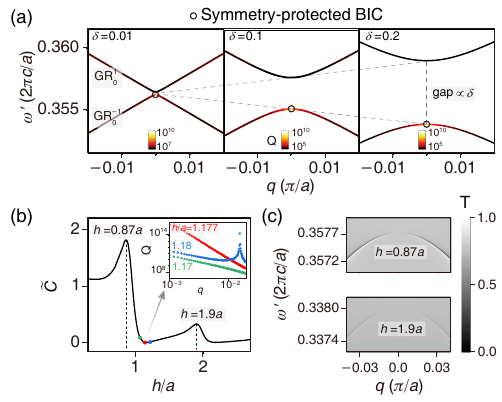}
    \caption{Symmetry protected BICs from degenerate guided mode resonances. (a) Dispersion of GR$_0^{-1}$ and GR$_0^1$ bands for $\delta=0.01$, 0.1, 0.2, and $h=a$. The bandgap scales linearly with $\delta$, and the BIC remains pinned at the $\Gamma$ point. (b) Proportionality constant $\tilde C$ in $\delta\omega''=\tilde C\,q^2\delta^2$ vs.\ slab thickness $h$. At $h=1.177a$, $\tilde C=0$ leads to a \textit{Q} scaling change from $q^{-2}$ to $q^{-6}$. The inset shows the \textit{Q} factors for three thicknesses. (c) Transmission spectra: for a small $\tilde{C}$ at $h=1.9a$, the resonance remains narrow over a wide $q$, whereas for a large $\tilde{C}$ at $h=0.87a$, it broadens significantly.
    }
    \label{fig:SP-BIC} 
\end{figure}

The degeneracy of $m = \pm1$ states at the $\Gamma$ point implies $f_{m,1} = f_{m,-1}$ and $f_{1,m} = f_{-1,m}$, and the same holds for their derivatives with respect to $\omega$ and $\delta$. The above determinant condition simplifies to the quadratic equation
\begin{equation}
a^{\prime\prime}\delta\omega^2 + b^{\prime\prime}\delta^2\delta\omega + (c_{1}^{\prime\prime}\delta^2 + c_{2}^{\prime\prime}\delta^3) = 0,
\end{equation}
where
\begin{equation*}
\begin{array}{lll}
a^{\prime\prime} &= \frac{1}{2}f_{00}(\partial_\omega f_{1,1})^2 = \mathrm{O}(1), \\
b^{\prime\prime} &= -\frac{1}{\sqrt{2}}(
u_{+,0}^2 + u_{-,0}^2)f_{01}f_{10}\partial_\omega f_{1,1} = \mathrm{O}(1), \\
c_1^{\prime\prime} &= -\frac{1}{2}f_{00}(\partial_\delta f_{1,1})^2 = \mathrm{O}(1), \\
c_2^{\prime\prime} &= \frac{1}{\sqrt{2}}(u_{+,0}^2 + u_{-,0}^2)f_{01}f_{10}\partial_\delta f_{1,1} = \mathrm{O}(1).
\end{array}
\end{equation*}
Solving the quadratic equation and expanding $\delta \omega$ in a Taylor series around $\delta = 0$ gives
\begin{equation*}
\delta\omega = \pm \sqrt{\frac{-c_1^{\prime\prime}}{a^{\prime\prime}}} \delta + \frac{-b^{\prime\prime} \pm a^{\prime\prime}c_2^{\prime\prime}/\sqrt{-a^{\prime\prime}c_1^{\prime\prime}}}{a^{\prime\prime}}\delta^2 + \mathrm{O}(\delta^3).
\label{eq:omega-at-Gamma-degeneratecase-1}
\end{equation*}
The first term, $\sqrt{-c_1^{\prime\prime}/a^{\prime\prime}} = \partial_\delta f_{1,1} / \partial_\omega f_{1,1} \in \mathbb{R}$, is real because $f_{11}$ is real-valued. Thus, the first-order correction to $\delta\omega$ is real, indicating an avoided level crossing and the opening of a bandgap of width $\mathrm{O}(\delta)$. The second-order correction vanishes for the negative root, whereas the positive root yields a complex value owing to the complex nature of $f_{00}$ and $f_{10}$. The former case implies the presence of a symmetry-protected BIC pinned at the $\Gamma$ point.
Noted that only one solution exhibits a vanishing imaginary part; the other remains leaky. Thus, unlike accidental or Friedrich--Wintgen BICs, symmetry-protected BICs lack a true dual counterpart. 
As illustrated in figure~\ref{fig:SP-BIC}(a), symmetry-protected BICs (marked by circles) are shown on the lower band for the perturbation strengths $\delta = 0.01$, $0.1$, and $0.2$, with slab thickness $h = a$. It is evident that both the bandgap width and frequency shift of the symmetry-protected BICs scale linearly with $\delta$, in excellent agreement with the numerical simulations.

A comparison with the analysis of the off-$\Gamma$ crossing point in equation~(\ref{eq:GRdisp-3band-cross2}) reveals similar characteristics. 
In both cases, the avoided crossing of the guided-mode resonances results in a bandgap scaling with $\delta$, regardless of whether the crossing occurs at the $\Gamma$ point. However, the behaviour of the $\delta^2$ term distinguishes the two types of crossings. Ordinary guided-mode resonances exhibit a nonzero $\delta^2$ term in the frequency shift, whereas the symmetry-protected BICs feature a vanishing $\delta^2$ correction. This contrast arises from the bulk Bloch wave $|\psi_0\rangle$, corresponding to the leaky FP modes, which significantly contributes to $\delta\omega''$ in the guided-mode resonances but plays no role in the symmetry-protected BICs.

In addition to the symmetry-protected BIC, the resonant states in its vicinity have attracted significant interest owing to their extremely high $Q$ factors~\cite{liu2019high, koshelev2018asymmetric, wu2019giant, sun2023infinite}. 
The conventional perturbation theory for non-degenerate states cannot be directly applied to the analysis of the resonant states near the $\Gamma$ point, even though these states are not exactly degenerate. Instead, a perturbative approach within a near-degenerate subspace must be employed. This approach resembles that for degenerate states, with the key modification involving the introduction of a small wavevector $q$. The resulting eigenvalues are determined as follows:
\begin{equation}
    \lambda_{\pm} = \overline{\varepsilon}k_0^2 - (q^2 + m^2 G^2) \pm \sqrt{(2qmG)^2 + k_0^4 |\langle -m | \varepsilon(x) | m \rangle |^2 \delta^2}.
    \label{eq:eigenvaluepertur-small-q}
\end{equation}
It is evident that the eigenvalues converge to those in the degenerate case shown in equation~(\ref{eq:eigen-sys-pertur-degeneracy}) as $q \to 0$. 
The corresponding eigenstates closely resemble their degenerate counterparts, except for the replacement of the original zero-order wave functions $|\psi_{\pm}^{(0)}\rangle = \frac{1}{\sqrt{2}} (|{-}m\rangle \pm |m\rangle)$ by
\[
|\psi_{\pm}^{(0)}\rangle = \cos\theta_{\pm} |{-}m\rangle + \sin\theta_{\pm} |m\rangle.
\]
Here, the phase angles 
\begin{equation*}
    \theta_{\pm} = \arctan \Big( \frac{k_0^2 \langle -m | \varepsilon(x) | m \rangle \delta}{2qmG \pm \sqrt{(2qmG)^2 + k_0^4 |\langle -m | \varepsilon(x) | m \rangle|^2 \delta^2}} \Big),
\end{equation*}
are $q$-dependent. As $q \to 0$, the phase angles approach $\theta_\pm \to \pm \pi/4$, and the zero-order wave functions recover their degenerate form. The condition for guided-mode resonances near the $\Gamma$ point is given by
\begin{equation}
    \mathrm{det}
    \left[
    \begin{array}{ccc}
        f_{00} & u_{-,0} f_{-1,0}\delta & u_{+,0} f_{1,0}\delta \\
        u_{0,-1} f_{0,-1}\delta & \cos\theta_{-} f_{-1,-1} & \cos\theta_{+} f_{1,-1} \\
        u_{0,1} f_{0,1}\delta & \sin\theta_{-} f_{-1,1} & \sin\theta_{+} f_{1,1}
    \end{array}
    \right] = 0.
    \label{eq:GR-3bands-qsmall}
\end{equation}

For an unperturbed waveguide mode characterised by $(q - G, \omega_0)$, the first-order correction $\delta\omega$ can be evaluated for any specified value of $q - G$. Considering the resonant modes on the lower band with a small wavevector $q > 0$, the matrix elements $f_{mn}$ ($m,n \neq 0$) are expanded via a Taylor series in two variables: $\delta\omega$ and $\delta$.
Notably, the eigenvalue perturbation in equation~(\ref{eq:eigenvaluepertur-small-q}) includes only the terms proportional to $\delta^2$, resulting in a Taylor expansion involving only even powers of $\delta$. This behaviour contrasts with the degenerate case, where a linear term in $\delta$ is present. Consequently, $f_{mn}$ ($m,n \neq 0$) can be expanded as
\begin{equation*}
    f_{mn}(\omega,\delta) = f_{mn}(\omega_0,0) + \delta\omega\, \partial_\omega f_{mn}(\omega_0,0) + \frac{1}{2} \delta^2\, \partial_\delta^2 f_{mn}(\omega_0,0) + \cdots.
\end{equation*}
When considering the band GR$_0^{-1}$, we have $f_{-1,-1}(\omega_0) = 0$, whereas $f_{1,-1}$ and $f_{11}$ remain nonzero on this band.
Substituting these into equation~(\ref{eq:GR-3bands-qsmall}), the leading-order term of the determinant becomes linear in $\delta\omega$. Therefore, $\delta\omega$ can be explicitly expressed as a polynomial in $\delta$, where the coefficient of $\delta^2$ corresponds to $C(q)$ defined in equation~(\ref{eq:C_q}). 
Again, we obtain $\delta\omega'' = \mathrm{Im}(C)\, \delta^2$ for guided-mode resonances near the $\Gamma$ point. For a small $q$, the proportionality constant can be further expanded as
\begin{equation*}
    \delta\omega'' = \tilde{C} \cdot q^2 \cdot \delta^2,
\end{equation*}
where $\tilde{C}$ is a function of the structural parameters. 

When $\tilde{C} \neq 0$, the expression given above accurately describes the imaginary part $\omega''$ of the resonant state near a symmetry-protected BIC. However, when $\tilde{C} = 0$, the estimation of $\omega''$ requires the inclusion of higher-order corrections in the perturbation theory. In fact, the condition $\tilde{C} = 0$ corresponds precisely to the coalescence of two accidental BICs with a symmetry-protected BIC. Near this point, as the accidental BIC approaches the symmetry-protected BIC, the imaginary part scales as $\delta \omega'' \propto q^2(q - q_*)^2(q + q_*)^2$.
As shown in figure~\ref{fig:SP-BIC}(b), $\tilde{C}$ is plotted as a function of the thickness $h$. Clearly, $\tilde{C}$ exhibits its extrema (including a zero value) at specific thicknesses.
The inset of figure~\ref{fig:SP-BIC}(b) illustrates the $Q$ factors of the resonant states near the $\Gamma$ point at different thicknesses $h$. Notably, at the critical thickness $h = 1.177a$, we have $\tilde{C}=0$, and the $Q$ factor scales as $q^{-6}$ when $q_* \to 0$, consistent with previous findings~\cite{Peng2019HighQgr}. 
For comparison, the transmission spectra at two other thicknesses ($h = 0.87a$ and $1.9a$) are shown in figure~\ref{fig:SP-BIC}(c). 
At $h = 1.9a$, the spectral peak remains narrow over a broad range of wavevectors, indicating a robust high-$Q$ resonance. In contrast, at $h = 0.87a$, the sharp peak is sustained only over a narrow range of $q$, consistent with the larger value of $\tilde{C}$. The proportionality constant $\mathrm{Im}(C) = \tilde{C} \cdot q^2$, derived from first principles, provides critical insight into the realisation of stable high-$Q$ resonances over a broad angle range.

\subsection{Interaction of guided-mode resonances with orthogonal polarisations: far-field radiation and EPs} \label{sec:GRs-TE&TM}

In the two- and three-band models discussed above, the bands involved share the same polarisation. From the perspective of the perturbation theory, the interaction between bands with a small energy spacing plays a significant role, particularly at the degenerate points of the energy bands.
As discussed at the beginning of section~\ref{sec:FW-BIC}, a three-band model is sufficient to capture the known features of the BICs along the high-symmetry lines of the BZ. The bulk Bloch waves forming these BICs possess the same polarisation; thus, degeneracies arising from polarisation do not occur.

In 2D structures, the energy bands form surfaces in the $(\textbf{k}_{||}, \omega)$ space. The intersection of two such bands typically results in a line, whereas three bands can intersect at a point. Intersections involving more than three bands are extremely rare. The number `three' in the three-band model originates from the co-dimension of a point (i.e. zero-dimensional). However, the situation becomes more complicated when the polarisation degree of freedom is introduced, as it leads to a new kind of degeneracy.

In the cases considered above, although Bloch waves with orthogonal polarisations do not interact along high-symmetry lines, they do interact away from these lines. This interaction is also the reason why the far-field radiation of the leaky mode cannot be treated as linearly polarised~\cite{zhen2014topological}. 
To incorporate polarisation effects, we consider vector waves instead of scalar waves from the outset. The formalism should therefore be adapted to include all polarisation states. EM waves, as transverse waves, exhibit a twofold degeneracy in isotropic and homogeneous media, corresponding to the so-called \textit{s}- and \textit{p}-polarisations. These two eigenstates are orthogonal and can be treated independently as scalar waves within their respective eigenspaces. However, such degeneracy is eliminated in periodic media.

In the 1D PhC considered above, the two polarisation states are referred to as E- and H-modes, where the respective $\textbf{E}$ and $\textbf{H}$ fields are perpendicular to the direction of periodicity. Because these modes are defined with respect to the layer interface (the $y$-$z$ plane in this case) rather than the plane of incidence (as in the \textit{s}- and \textit{p}-waves), a change in basis is required to account for both the uniform background medium and periodic structure when analysing PhC slabs.

The electric field of an E-mode and magnetic field of an H-mode can be written as $(0, E_{\mathrm{e}y}, E_{\mathrm{e}z})$ and $(0, H_{\mathrm{h}y}, H_{\mathrm{h}z})$, respectively. Given that the current system is uniform in the $y$-direction and periodic in the $x$-direction, with the periodic modulation given by equation~(\ref{eq:epperiodic}), we can fix the $k_y$ component of the wavevector for an eigenstate, whereas $k_z$ is determined from the wave equation.
Using perturbation theory, we obtain the field components of the Bloch modes---namely, the electric field for the E-mode and the magnetic field for the H-mode (see supplementary section 4 for the detailed derivation). The corresponding magnetic field for E-mode and electric field for H-mode then follow directly from Maxwell's equations. This approach yields two families of Bloch modes with mutually orthogonal polarisations, illustrated schematically in figure 7(a), which together form a complete basis for describing EM fields in the 1D PhC slab.

\begin{figure}[h]
    \centering    
    \includegraphics[width=0.8\textwidth]{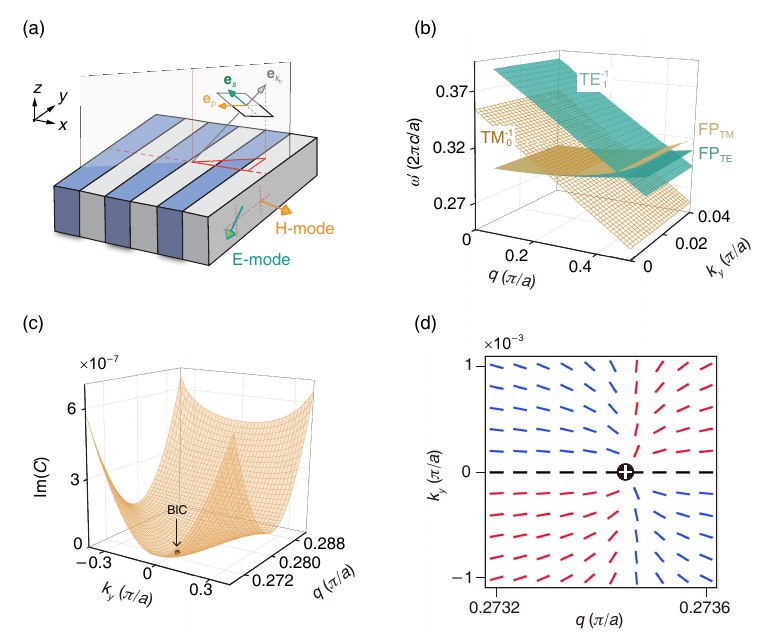}
    \caption{Interaction between different polarisation states off the high-symmetry line. (a) Definition of outgoing polarisations $\mathbf{e}_s=(k_y,-q_n,0)/k_\parallel$ and $\mathbf{e}_p=\mathbf{e}_k\times\mathbf{e}_s$, and the polarisations of E  and H modes of bulk Bloch waves. (b) Four band model combining two polarisations: guided modes TE$_1^{-1}$, TM$_0^{-1}$, and their FP counterparts. (c) Calculated Im$(C)$ for TM$_0^{-1}$ showing a zero point (fixed point under perturbation). (d) Far field polarisation states (red: right-handed, blue: left-handed) at $\delta=0.01$; the BIC is a polarisation singularity with nonzero topological charge. Slab thickness $h=1.67a$. 
    }
    \label{fig:polarization-off-kx-axis} 
\end{figure}
 
To properly account for the radiation condition at $z = \pm h/2$, the fields outside the PhC slab are typically decomposed into \textit{s}- and \textit{p}-polarised waves. The plane of incidence is defined by the wavevector $\textbf{k}_\mathrm{b}$ and unit vector $\textbf{e}_z$. The two orthonormal polarisation vectors are defined as $\textbf{e}_s = (k_y, -q_n, 0)/k_{||}$ and $\textbf{e}_p = \textbf{e}_{k_\mathrm{b}} \times \textbf{e}_s$, which, together with $\textbf{e}_{k_\mathrm{b}}$, form a right-handed coordinate frame. To describe the field components in the $x$-$y$ plane, we further define the in-plane wavevector $\textbf{k}_{||n} = (q_n, k_y, 0)$ and the projection of $\textbf{e}_p$ onto this plane as $\textbf{e}_{p,||} = (q_n, k_y, 0)/k_{||n}$. When focusing on the polarisation singularities of far-field radiation, it is convenient to use $\textbf{e}_{p,||}$ directly rather than rotating $\textbf{e}_p$ into the $x$-$y$ plane~\cite{yoda2020generation}.

Inside the slab, the electric and magnetic fields are expressed as superpositions of the E- and H-mode fields given in the supplementary section 4.
\begin{equation}
\begin{array}{ll}
    \textbf{E}_\mathrm{ins} &= \sum_{n} \left( c_{\mathrm{e},n} \textbf{E}_{\mathrm{e},n} + c_{\mathrm{h},n} \textbf{E}_{\mathrm{h},n} \right), \\[5pt]
    \textbf{H}_\mathrm{ins} &= \sum_{n} \left( c_{\mathrm{e},n} \textbf{H}_{\mathrm{e},n} + c_{\mathrm{h},n} \textbf{H}_{\mathrm{h},n} \right).
\end{array}
\label{ep:EH-slab-ins}
\end{equation}
\noindent Outside the slab, the fields are expanded as superpositions of the outgoing \textit{s}- and \textit{p}-polarised plane waves
\begin{equation}
\begin{array}{ll}
    \textbf{E}_\mathrm{b} &= \sum_{j} \left( t_{s,j} \textbf{E}_{s,j}^\mathrm{b} + t_{p,j} \textbf{E}_{p,j}^\mathrm{b} \right), \\[5pt]
    \textbf{H}_\mathrm{b} &= \sum_{j} \left( t_{s,j} \textbf{H}_{s,j}^\mathrm{b} + t_{p,j} \textbf{H}_{p,j}^\mathrm{b} \right).
\end{array}
\label{ep:EH-slab-out}
\end{equation}

As shown in section~\ref{subsec: interaction of FP & GMR}, obtaining the correct value of $\delta\omega''$ requires the inclusion of at least two Bloch modes, such as those indexed by $0$ and $-1$. Given the two polarisations under consideration, the total number of Bloch states doubles: $(\mathrm{e,h}) \times (0, -1)$ inside the slab, and $(s,p) \times (0, -1)$ outside. The boundary conditions at $z = \pm h/2$ enforce the continuity of the tangential components of the electric and magnetic fields, as follows:
\begin{equation}
\begin{array}{ll}
    \sum\limits_{i = 0, {-1}}
    \left( c_{\mathrm{e},i} \textbf{E}_{\mathrm{e},i} + c_{\mathrm{h},i} \textbf{E}_{\mathrm{h},i} \right)_{\parallel} &=
    \sum\limits_{j = 0, {-1}} \left( t_{s,j} \textbf{E}_{s,j}^\mathrm{b} + t_{p,j} \textbf{E}_{p,j}^\mathrm{b} \right)_{\parallel}, \\[5pt]
    \sum\limits_{i = 0, {-1}}
    \left( c_{\mathrm{e},i} \textbf{H}_{\mathrm{e},i} + c_{\mathrm{h},i} \textbf{H}_{\mathrm{h},i} \right)_{\parallel} &=
    \sum\limits_{j = 0, {-1}} \left( t_{s,j} \textbf{H}_{s,j}^\mathrm{b} + t_{p,j} \textbf{H}_{p,j}^\mathrm{b} \right)_{\parallel},
\end{array}
\end{equation}
where the subscript $\parallel$ denotes the components parallel to the interface at $z = \pm h/2$.

Among the four vectors $\textbf{E}_{\mathrm{e},0}$, $\textbf{E}_{\mathrm{e},-1}$, $\textbf{E}_{\mathrm{h},0}$, and $\textbf{E}_{\mathrm{h},-1}$, none are generally mutually parallel. In fact, they are linearly independent at off-high-symmetry lines (i.e. for $k_y \neq 0$). Additionally, the $x$- and $y$-components of $\textbf{E}_0$ and $\textbf{E}_{-1}$ differ in phase. As a result, a $4 \times 4$ matrix system should be constructed to enforce the boundary conditions, rather than the simpler $2 \times 2$ matrix used in equation~(\ref{eq:GRdisp-2modes}). 
The far-field radiation emitted from the slab generally exhibits elliptical polarisation, involving both \textit{s}- and \textit{p}-components. Below, we first analyse the diagonalisation of the $4 \times 4$ matrix in the unperturbed case $\delta = 0$, where the far-field polarisation remains linear.

In the absence of perturbation, the only task is to transform the basis from the E- and H-modes inside the slab to the $s$- and $p$-waves outside. The electric field inside the slab can always be expressed as $c_{\mathrm{e}}\mathbf{E}_{\mathrm{e}} + c_{\mathrm{h}}\mathbf{E}_{\mathrm{h}}$. Thus, the $s$- and $p$-waves correspond to the linear combinations $(c_{s,{\mathrm{e}}}, c_{s,{\mathrm{h}}})$ and $(c_{p,{\mathrm{e}}}, c_{p,{\mathrm{h}}})$, respectively. Therefore, a transformation matrix $\Lambda$ can be defined to relate these two bases as
\begin{equation}
    \Lambda = 
    \left[
    \begin{array}{cc}
     c_{s,\mathrm{e}} & c_{s,\mathrm{h}} \\
     c_{p,\mathrm{e}} & c_{p,\mathrm{h}} \\
    \end{array}
    \right]
     = 
    \left[
    \begin{array}{cc}
        -q_{n}k_{zn} & k_{y}k_{zn} \\
        -k_{y}k_{0}^{2}/k_{zn} & -q_{n}k_{zn} \\
    \end{array}
    \right].
\end{equation}  

After eliminating the transmission coefficients $t_{s}$ and $t_{p}$ outside the slab, the boundary conditions can be formulated as a system of linear equations with the four variables $\{c_{\rm{e}0}, c_{\rm{h}0}, c_{\rm{e},{-1}}, and c_{\rm{h},{-1}}\}$. Applying the block-diagonal transformation matrix $\Lambda_0 \oplus \Lambda_{-1}$ changes the basis to $\{c_{s0}, c_{p0}, c_{s,{-1}}, and c_{p,{-1}}\}$. Then, the guided-mode resonance condition takes the following form:

\begin{equation}
    \det
    \left[
    \begin{array}{cc}  
        \begin{array}{cc}
            f_{s_{0},s_{0}} & \mathrm{O}(\delta^2) \\
            \mathrm{O}(\delta^2) & f_{p_{0},p_{0}}
        \end{array}
        &
        \mathrm{O}(\delta)_{2\times2}
        \\  
         \mathrm{O}(\delta)_{2\times2}
        &
        \begin{array}{cc}
            f_{s_{{-1}},s_{{-1}}} & \mathrm{O}(\delta^2) \\
            \mathrm{O}(\delta^2) & f_{p_{{-1}},p_{{-1}}}
        \end{array}
    \end{array}
    \right]
    = 0.
\label{eq:GRdisp-4modes}
\end{equation}

For brevity, we present only the order of the matrix elements with respect to $\delta$, omitting their explicit expressions owing to their complicated forms; here $\mathcal O(\delta)_{2\times2}$ denotes a generic $2\times2$ matrix whose elements are all of order $\delta$. We note that the vanishing of the diagonal terms in this $4 \times 4$ matrix yields the dispersion relations for the unperturbed states ($\delta=0$), such as FP modes and waveguide modes. Assume that the $E_{\mathrm{e}y}$ component of the E-mode is odd; then, the replacement can be formulated as $c_{n}e^{ik_{zn}z} + d_{n}e^{-ik_{zn}z} \to c_{n}\sin(k_{zn}z)$. The impedance conditions for \textit{s}- and \textit{p}-waves are
\begin{equation}
    \begin{array}{cc}
        k_{zn}\cot(k_{zn}h/2) = ik_{zn}^{\mathrm{b}}, \\[5pt]
        \frac{k_{zn}}{\overline{\epsilon}}\tan(k_{zn}h/2) = i\frac{k_{zn}^{\mathrm{b}}}{\epsilon_{\mathrm{b}}}.
    \end{array}
\end{equation}
These expressions are special cases of an equation~(\ref{eq:wgm-DtN}) for odd modes with wavevector $q - nG$. 
We note that the resonant modes in a PhC slab can also be classified according to their mirror symmetry with respect to the $x$-$y$ plane~\cite{joan08}. 
For TE-like modes, the electric field is predominantly parallel to the mirror plane, whereas for TM-like modes, it is primarily perpendicular. In the far field, TE-like and TM-like modes correspond to $s$- and $p$-polarised waves, respectively, provided that the coupling between them is weak.

When a PhC slab with a nonzero perturbation ($\delta \neq 0$) is considered, the unperturbed waveguide mode undergoes a frequency correction $\omega = \omega_{0} + \delta\omega$. For the $\mathrm{TM}_{0}^{-1}$ band, the correction to the matrix element $f_{p_{{-1}},p_{{-1}}}$ is given by $\delta\omega \, \partial_{\omega} f_{p_{{-1}},p_{{-1}}}$, evaluated at the unperturbed waveguide condition $f_{p_{{-1}},p_{{-1}}}(q-G, k_y, \omega_0) = 0$. It can be demonstrated from the $4\times4$ matrix structure that $\delta\omega$ exhibits a quadratic dependence on the perturbation strength, i.e. $\delta\omega \propto \delta^2$, where the proportionality constant is denoted by $C(q, k_y)$, similar to that in equation~(\ref{eq:C_q}). Accordingly, we obtain
\begin{equation*}
    \delta\omega^{\prime\prime} = \mathrm{Im}(C)\delta^2,
\end{equation*}
for the guided-mode resonances. As an example, figure~\ref{fig:polarization-off-kx-axis}(b) shows the four relevant bands: the guided-mode resonances TE$_1^{-1}$ and TM$_0^{-1}$, along with two FP modes denoted by $\mathrm{FP}_{\mathrm{TE},3}$ and $\mathrm{FP}_{\mathrm{TM},2}$. The subscripts of the FP modes indicate the number of field nodes in the electric and magnetic fields along the $z$-direction. Away from the high-symmetry line, all four resonant modes can couple. However, along the high-symmetry line, coupling occurs only between modes with the same polarisation. As shown in figure~\ref{fig:polarization-off-kx-axis}(c), a theoretical map of $\mathrm{Im}(C)$ for the TM$_0^{-1}$ band is plotted in the momentum space. A fixed point with $\mathrm{Im}(C) = 0$ is identified, indicating that as $\delta \to 0$, the BIC converges to this point. For a fixed $q$, $\mathrm{Im}(C)$ is an even function of $k_y$, owing to the mirror symmetry of the system along the $y$-axis.

In addition to the band structure and its corresponding imaginary parts, our theoretical framework can be directly applied to determine the far-field polarisation of the resonant states. By substituting the complex resonant frequency $\omega$ into the $S$-matrix in equation~(\ref{eq:S-poles}), a set of nontrivial solutions for the coefficients $\{ c_{s0}, c_{p0}, c_{s,{-1}}, c_{p,{-1}} \}$ can be obtained. These solutions can further provide the transmission coefficients $\{ t_{s0}, t_{p0}, t_{s,{-1}}, t_{p,{-1}} \}$ outside the slab through the radiation boundary condition.

For the $\mathrm{GR}_{0}^{-1}$ band, the far-field polarisation is characterised by the vector field $\{ t_{s0}, t_{p0} \}$. Figure~7(d) presents the far-field polarisation pattern of the $\mathrm{TM}_{0}^{-1}$ band. A BIC exhibits no outgoing radiation, rendering its far-field polarisation undefined. This gives rise to a polarisation vortex, centred at the BIC, also referred to as a polarisation singularity with a nonzero winding number. The predicted far-field polarisation vortices are further validated by full-wave numerical simulations, showing excellent agreement in both polarisation states and topological charge (see supplementary section 4).

\begin{figure}[h]
    \centering    
    \includegraphics[width=0.8\textwidth]{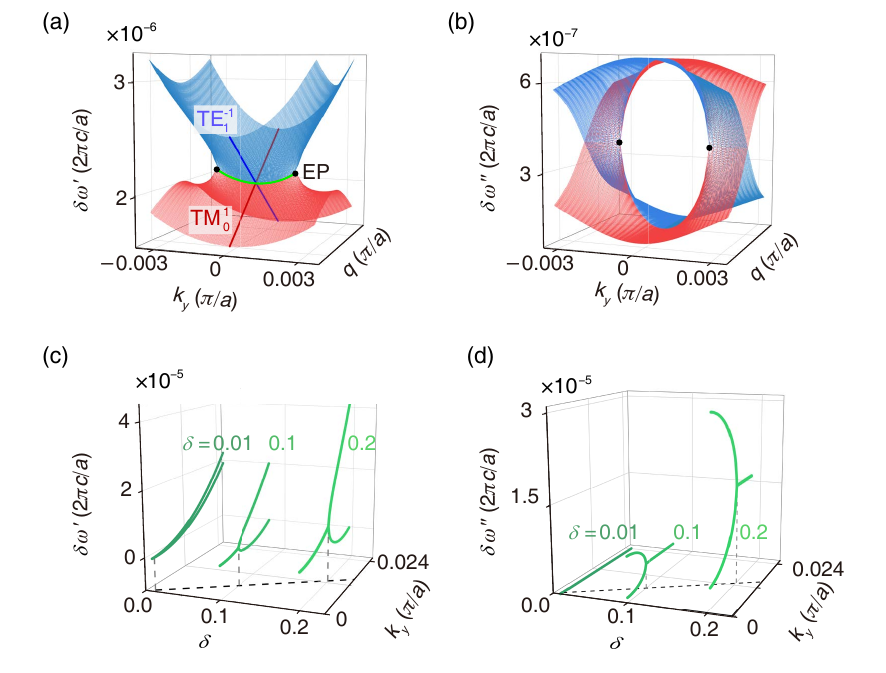}
    \caption{Exceptional points (EPs) from the interaction between bands with different polarisations. (a) Real and (b) imaginary parts of the band structure for $h=a$ and perturbation $\delta=0.03$; EPs (black dots) connected by a Fermi arc (green). (c, d) Intersection lines of the real and imaginary parts as a function of $k_y$ for various $\delta$, illustrating the linear shift of the EPs with $\delta$.
    }
    \label{fig:EPs} 
\end{figure}

In addition to polarisation singularities, another class of band singularities, known as EPs, can also be interpreted using a model that incorporates the interactions among multiple bands. EPs arise when two or more eigenmodes coalesce into a single mode owing to non-Hermitian effects. This phenomenon is not restricted to $\mathcal{PT}$-symmetric systems with balanced gain and loss but also occurs in systems exhibiting only radiative loss. The differential radiation losses of the resonant modes play a critical role in the formation of EPs~\cite{wang2024topological}.

At least two resonant states are required for the generation of EPs. Together with FP modes, we should consider three Bloch states, for example, those indexed by $0$, $-1$, and $1$. In comparison with the case of conventional Friedrich--Wintgen BICs discussed in section~\ref{sec:FW-BIC}, the total number of Bloch states is doubled, that is, $(\mathrm{e}, \mathrm{h}) \times (0, -1, 1)$ and $(s, p) \times (0, -1, 1)$, for the regions inside and outside the slab, respectively. The boundary conditions connecting the $\mathbf{E}$ and $\mathbf{H}$ fields inside and outside the slab, as given in equations~(\ref{ep:EH-slab-ins}) and (\ref{ep:EH-slab-out}), yield a system of linear equations involving six variables: $\{ c_{s0}, c_{p0}, c_{s,-1}, c_{p,-1}, c_{s1}, c_{p1} \}$. Then, the guided-mode resonance condition is given by:
\begin{equation}
    \mathrm{det}
    \left[
        \begin{array}{ccc}
            \begin{array}{cc}
               f_{s_{0},s_{0}} & \mathrm{O}(\delta^2) \\
               \mathrm{O}(\delta^2) & f_{p_{0},p_{0}} \\
            \end{array}
            &  \mathrm{O}(\delta)_{2\times2} 
            &  \mathrm{O}(\delta)_{2\times2} \\
             \mathrm{O}(\delta)_{2\times2} &
            \begin{array}{cc}
                f_{s_{{-1}}, s_{{-1}}} & \mathrm{O}(\delta^2) \\
                \mathrm{O}(\delta^2) & f_{p_{{-1}}, p_{{-1}}} \\
            \end{array}
            &  \mathrm{O}(\delta)_{2\times2} \\
             \mathrm{O}(\delta)_{2\times2} &  \mathrm{O}(\delta)_{2\times2} & 
            \begin{array}{cc}
               f_{s_{1},s_{1}} & \mathrm{O}(\delta^2) \\
               \mathrm{O}(\delta^2) & f_{p_{1},p_{1}} \\
            \end{array}
        \end{array}
    \right] = 0. 
    \label{eq:GRdisp-6modes}
\end{equation}

We take two guided-mode resonances orthogonal along the $k_x$-axis—$\mathrm{TE}_1^{-1}$ and $\mathrm{TM}_0^1$—as an illustrative example. The real parts of their complex frequencies exhibit degeneracy at the crossing point $(q_c, 0; \omega_c)$ owing to their orthogonal polarisations, independent of the perturbation $\delta$. As $k_y$ increases, the perturbation becomes involved, and these resonances interact, potentially giving rise to EPs near the crossing point. At the crossing point, the conditions $f_{s_{-1}, s_{-1}} = 0$, and $f_{p_{1}, p_{1}} = 0$ are satisfied. We perform a Taylor expansion around this point, as follows:
\begin{equation*}
    \begin{array}{ll}
        f_{s_{-1}, s_{-1}} &= \left(\delta\omega\,\partial_{\omega} + \frac{1}{2}\delta^2\,\partial_{\delta}^{2} + \delta q\,\partial_{q} + \frac{1}{2}k_{y}^{2}\,\partial_{k_{y}}^2\right) f_{s_{-1}, s_{-1}} + \cdots, \\
        f_{p_{1}, p_{1}} &= \left(\delta\omega\,\partial_{\omega} + \frac{1}{2}\delta^2\,\partial_{\delta}^{2} + \delta q\,\partial_{q} + \frac{1}{2}k_{y}^{2}\,\partial_{k_{y}}^2\right) f_{p_{1}, p_{1}} + \cdots.
    \end{array}
\end{equation*}

\noindent By substituting these expansions into equation~(\ref{eq:GRdisp-6modes}), the leading-order terms yield a quadratic equation in $\delta\omega$. The condition for an EP is that this equation has a degenerate root of multiplicity two, ensuring coalescence of the upper and lower band states. This enables the determination of the location of EPs in the momentum space and their evolution with the perturbation strength $\delta$.

As illustrated in figure~\ref{fig:EPs}(a) and (b), the band structure and corresponding imaginary parts of $\mathrm{TE}_{1}^{-1}$ and $\mathrm{TM}_{0}^{1}$ are computed for a structure with $h = a$ and perturbation strength $\delta = 0.03$. A pair of EPs (black dots) is connected by a Fermi arc, represented by a green line, in the $(\mathbf{k}_{||},\omega)$ space, along which the real parts of the complex bands remain degenerate. The loci in the momentum space, where either the real or imaginary parts of the eigenfrequencies of the two interacting bands become equal, are extracted and plotted in figure~\ref{fig:EPs}(c) and (d). The evolution of the EP positions under variation in the perturbation strength $\delta$ is indicated by the black dashed line. It is evident that the EPs originate at the intersection of the TE$_1^{-1}$ and TM$_0^1$ bands along the $k_x$-axis and shift away from this axis proportionally to $\delta$. This perturbative model not only accounts for the generation of EPs but also predicts their evolution in the momentum space with increase in $\delta$.

\subsection{BICs in 2D PhC slabs}
\label{sec:BIC-2D}

For 1D PhC slabs, we employ a perturbation theory based on the $S$-matrix to investigate the optical modes, yielding both the complex band structure and far-field polarisation. This framework enables the identification of various types of BICs, as well as band singularities such as EPs. A two-band model is sufficient to describe accidental BICs, whereas a three-band model is required to accurately predict Friedrich--Wintgen and symmetry-protected BICs. Notably, the interactions between bands with closely spaced energy levels are crucial. Thus, the minimal dimension of the Hilbert space can be determined, which does not exceed the number of propagating Bloch waves involved, as shown in equation~(\ref{eq:S-min-rank}).

Although the above results focus on 1D PhC slabs, our approach is universal and readily extends to 2D periodic systems. In a 2D PhC slab, discrete translational symmetry exists in the $x$-$y$ plane. As discussed in the reference~\cite{joan08}, guided-mode resonances can also be classified as either TE-like or TM-like modes with respect to the $x$-$y$ mirror plane. Analogous to the 1D case, for the magnetic field in this context, the \textbf{H} field is predominantly perpendicular to the mirror plane for TE-like modes, whereas it is predominantly parallel for TM-like modes. In contrast to 1D PhC slabs, where the TE and TM modes remain orthogonal along high-symmetry lines, the 2D periodicity allows TE-like and TM-like modes to possess the same spatial symmetry and interact with each other, leading to a richer physical behaviour.

To determine the minimal Hilbert space required to estimate the imaginary part of the frequency, $\omega^{\prime\prime}$, it is necessary to identify the propagating Bloch waves that must be considered in a 2D PhC slab. In the 1D case with periodicity along the $x$-direction, at least two Bloch waves are required to capture the essential physics. In contrast, for 2D PhC slabs, which exhibit symmetry in the $x$-$y$ plane, band folding in the 2D reciprocal space should be considered. Consequently, the Hilbert space in 2D may have a higher dimension owing to the additional Bloch waves arising from this band folding. When polarisation is considered (i.e. distinguishing between TE-like and TM-like modes), the number of required Bloch waves further doubles. Thus, the first step is to determine the minimal Hilbert space dimension necessary to capture the essential physics of the 2D PhC slab based on a rigorous symmetry analysis of the vector field.

Consider a 2D PhC slab with lattice vectors $\textbf{a}_1$ and $\textbf{a}_2$ in the $x$-$y$ plane. The dielectric function is given by $\epsilon(\mathbf{r}) = \overline{\epsilon} + \varepsilon(\textbf{r}_{||})\delta$, where the periodic modulation $\varepsilon(\textbf{r}_{||}) = \varepsilon(\textbf{r}_{||}+\textbf{a}_i)$ is of the order $\mathrm{O}(1)$, and the small parameter $\delta$ is explicitly singled out. The perturbation theory applied here is analogous to that used for the 1D PhC slab. For the non-magnetic system considered here, the divergence-free magnetic field is $\mathbf{H} = (H_x, H_y, H_z)$. The governing equation for $\mathbf{H}$, equation~(\ref{eq:masterEq-H}), can be rewritten as:
\begin{equation}
    \nabla^2\mathbf{H} + \epsilon(\mathbf{r})k_{0}^{2}\mathbf{H} + \frac{1}{\epsilon(\mathbf{r})}\nabla\epsilon(\mathbf{r})\times(\nabla\times\mathbf{H}) = 0.
\end{equation}
For a given eigenstate, we fix the wavevector components $(k_x, k_y)$, whereas $k_z$ is determined by solving the wave equation. For a nonzero perturbation ($\delta \neq 0$), we analyse the perturbed eigenvalue problem for a general vector field $\psi$, which is $\mathbf{H}$ in this context, but can also represent $\mathbf{E}$ in other cases. The equation takes the form:
\begin{equation}
    ((\partial_{x}^{2} + \partial_{y}^{2} + \overline{\epsilon}k_{0}^{2})\mathbf{I} + \mathcal{V}\delta) \psi = k_{z}^{2} \psi,
\end{equation}
where $\mathcal{V}$ is the perturbation operator defined as:
\begin{equation*}
    \mathcal{V} = \varepsilon(\textbf{r}_{||})k_{0}^{2}\mathbf{I} + \frac{1}{\epsilon(\textbf{r})}
    \left[
    \begin{array}{ccc}
        -\frac{\partial\varepsilon}{\partial y}\partial_y & \frac{\partial\varepsilon}{\partial y}\partial_x & 0 \\
        \frac{\partial\varepsilon}{\partial x}\partial_y & -\frac{\partial\varepsilon}{\partial x}\partial_x & 0 \\
        \frac{\partial\varepsilon}{\partial x}\partial_z & \frac{\partial\varepsilon}{\partial y}\partial_z & -\frac{\partial\varepsilon}{\partial x}\partial_x - \frac{\partial\varepsilon}{\partial y}\partial_y
    \end{array}
    \right].
\end{equation*}

\noindent Unlike the 1D PhC slab, where the states, whether on or off high-symmetry lines, can be treated as scalar fields (either \textbf{E} or \textbf{H}), the 2D PhC slab requires a full vectorial treatment. As a first example, consider a square lattice with lattice vectors $\textbf{a}_1 = a \textbf{e}_x$ and $\textbf{a}_2 = a \textbf{e}_y$. Owing to the divergence-free condition of $\mathbf{H}$, only two field components are linearly independent. These can be described using the following orthonormal basis in the unperturbed case ($\delta = 0$):
\begin{equation}
    \begin{array}{l}
        |mn;1\rangle = \frac{1}{N_1}e^{i(k_{xm}x + k_{yn}y + k_{z}z)} (k_{z}, 0, -k_{xm})^{\mathrm{T}}, \\[4pt]
        |mn;2\rangle = \frac{1}{N_2}e^{i(k_{xm}x + k_{yn}y + k_{z}z)} (k_{xm}k_{yn}, -k_{xm}^{2}-k_{z}^{2}, k_{z}k_{yn})^{\mathrm{T}},
    \end{array}
    \label{eq:planewaves-2D}
\end{equation}
\noindent where $N_1$ and $N_2$ are normalisation constants. The $k_{z}$ component satisfies $k_{z}^{2} = \overline{\epsilon}k_{0}^{2} - k_{xm}^{2} - k_{yn}^{2}$, with $k_{xm}=k_x+mG$ and $k_{yn}=k_{y}+nG$, where $m$ and $n$ are the band-folding indices along the $x$- and $y$-directions, respectively. The perturbation theory then yields:
\begin{equation}
    \begin{array}{l}
        \lambda_{mn;\sigma} = \overline{\epsilon}k_{0}^{2} - k_{xm}^{2} - k_{yn}^{2} + \mathrm{O}(\delta^{2}), \\[4pt]
        |\psi_{mn;\sigma}\rangle = |mn;\sigma\rangle + \sum\limits_{(m^{\prime}n^{\prime};\sigma') \neq (mn;\sigma)}\frac{\langle m^{\prime}n^{\prime};\sigma'| \mathcal{V} |mn;\sigma\rangle}{(k^2_{xm'} + k_{yn^{\prime}}^{2}) - (k^2_{xm} + k_{yn}^{2})} |m^{\prime}n^{\prime};\sigma'\rangle \delta.
    \end{array}
    \label{eq: perturbed lambda for 2D}
\end{equation}
\noindent For convenience, we define the matrix elements of $v_{mn\sigma;m'n'\sigma'}$ as follows:
\begin{equation*}
    v_{mn\sigma;m'n'\sigma'} = 
    \left\{
    \begin{array}{cl}
        \frac{\langle m'n';\sigma'| \mathcal{V} |mn;\sigma\rangle}{(k^2_{xm^{\prime}} + k_{yn^{\prime}}^{2}) - (k^2_{xm} + k_{yn}^{2})}, & \mathrm{for } \; (m^{\prime},n^{\prime}) \neq (m,n), \\[8pt]
        0, & \mathrm{for } \; (m^{\prime},n^{\prime}) = (m,n).
    \end{array}
    \right.
\end{equation*}

If the states under consideration lie along the $k_x$-axis, we can adopt the two-step perturbation approach outlined in section. However, it is necessary to examine whether two bulk Bloch waves are sufficient to obtain the imaginary part of the frequency, $\omega''$, as in the 1D case. This is because, for instance, in addition to the two states $|00;\sigma\rangle$ and $|{-}1,0;\sigma\rangle$, the states $|0,\pm 1;\sigma\rangle$ should also be considered. Therefore, a total of \textit{eight} states are involved, considering two polarisations and band folding from both the $x$- and $y$-directions.

These eight states can be decomposed into orthogonal subspaces according to the mirror symmetry $\sigma_v(xz)$, defined with respect to the fixed electric-field component $E_z$, as the following analysis is restricted to the $k_x$-axis with $k_y = 0$. For simplicity, $\sigma_v(xz)$ is henceforth denoted as $\sigma_v$, with the $x$-$z$ mirror plane implied. For $n = 0$ (i.e. no band folding along the $y$-direction), the basis vectors in equation~(\ref{eq:planewaves-2D}) with $\sigma {=} 1, 2$ are orthogonal and possess opposite mirror symmetries with eigenvalues $\sigma_{v} {=} \mp 1$, respectively. The corresponding four states can be written explicitly as $|0,0;\sigma{=}1,\sigma_v{=}{-1}\rangle$, $|0,0;\sigma{=}2,\sigma_v{=}1\rangle$, $|{-1},0;\sigma{=}1,\sigma_v{=}{-1}\rangle$, and $|{-1},0;\sigma{=}2,\sigma_v{=}1\rangle$. For $n = \pm 1$, corresponding to band folding along the $y$-direction, the states $|0,\pm1;\sigma{=}1,2\rangle$ are degenerate in frequency, and a perturbation theory for the degenerate case must be applied, similar to that discussed in section~\ref{sec:FW-BIC}. These four states can also be split into two subspaces according to $\sigma_{v} = \pm 1$.

Therefore, a basis transformation leads from the original states $|0, \pm 1; \sigma\rangle$ to the new basis states $|0, \pm; \sigma, \sigma_{v}\rangle$, where $\pm 1$ in the former denotes band folding along the $y$-direction with wavevector shifts of $\pm G$, whereas $\pm$ in the latter represents linear combinations of the original states. The four new basis states are given by:
\begin{equation*}
    \begin{array}{l}
        |0, \pm; \sigma {=} 1, \sigma_{v} {=} 1\rangle = \frac{1}{\sqrt{2}} (|0, -1; \sigma {=} 1\rangle - |0, 1; \sigma {=} 1\rangle), \\
        |0, \pm; \sigma {=} 1, \sigma_{v} {=} -1\rangle = \frac{1}{\sqrt{2}} (|0, -1; \sigma {=} 1\rangle + |0, 1; \sigma {=} 1\rangle), \\
        |0, \pm; \sigma {=} 2, \sigma_{v} {=} -1\rangle = \frac{1}{\sqrt{2}} (|0, -1; \sigma {=} 2\rangle - |0, 1; \sigma {=} 2\rangle), \\
        |0, \pm; \sigma {=} 2, \sigma_{v} {=} 1\rangle = \frac{1}{\sqrt{2}} (|0, -1; \sigma {=} 2\rangle + |0, 1; \sigma {=} 2\rangle).
    \end{array}
\end{equation*}
A detailed derivation of the eigenvalues and eigenstates in this four-fold degenerate case is provided in supplementary section 5.  

Because the eight basis states can be classified by their mirror symmetry $\sigma_v$, they can be partitioned into two subspaces corresponding to $\sigma_v = \pm 1$. Based on this symmetry classification, we conclude that the minimal Hilbert space must satisfy $2 \leq \mathrm{dim}(S) \leq 4$. However, four-band intersections rarely occur away from the $\Gamma$ point and typically arise only under special spatial symmetries. Therefore, in most band-crossing scenarios, a three-band model is sufficient to capture the essential physics. We will address the four-band case in the context of an accidentally degenerate point in a 2D PhC slab later.

Let us first examine whether a two-band model can adequately capture the essential physics of 2D PhC slabs, analogous to the 1D case. In the first step of the perturbation theory, we need to identify the basis states that couple under the perturbation $\delta \neq 0$. As discussed above, four of the basis states can couple if they share the same $\sigma_v$ symmetry: two states without folding in the $y$-direction---$| 0, 0; \sigma \rangle$ and $| {-}1, 0; \sigma \rangle$---and two states resulting from folding in the $y$-direction---$| 0, \pm; \sigma {=} 1 \rangle$ and $| 0, \pm; \sigma {=} 2 \rangle$. For simplicity, the $\sigma_v$ label is omitted, as all involved states share the same mirror symmetry.

We note that the interaction between $|{-}1,0;\sigma\rangle$ and $|0,\pm;\sigma{=}1,2\rangle$ does not contribute to the leading order of $\omega''$, as the latter is folded from outside the light cone and does not directly induce energy leakage when only one radiation channel is present. It has been demonstrated that the imaginary part $\omega''$ of the perturbed waveguide mode primarily arises from its interaction with the FP modes, dominated by the basis state $|00;\sigma\rangle$, as in the 1D case discussed in section. Consequently, the minimal Hilbert space required to describe the complex band structure remains two-dimensional, involving only $|00;\sigma\rangle$ and $|{-}1,0;\sigma\rangle$. The perturbed bulk Bloch wavefunctions, as given in equation~(\ref{eq: perturbed lambda for 2D}), are thus simplified as:
\begin{equation}
    \begin{array}{l}
        |\psi_{00;\sigma}\rangle = |00;\sigma\rangle + v_{00,\sigma;-1,0,\sigma}|{-}1,0;\sigma\rangle\delta + \cdots, \\
        |\psi_{-1,0;\sigma}\rangle = |{-}1,0;\sigma\rangle + v_{-1,0,\sigma;00,\sigma}|00;\sigma\rangle\delta + \cdots,
    \end{array}
\end{equation}
which lead to the two-band model in 2D PhC slabs. As an example, consider a TM mode along the $k_x$-axis ($k_y {=} 0$). Only the $y$-component of the \textbf{H} field is nonzero, corresponding to the mirror symmetry $\sigma_{v} {=} 1$ and polarisation $\sigma {=} 2$. The vector field $\psi$ (which refers to the \textbf{H} field here) can be written as $|\psi_{-1,0;\sigma=2}\rangle = (0, H_y, 0)^{\mathrm{T}}$. The corresponding \textbf{E} field is given by $\frac{i}{\omega\epsilon(\mathbf{r})}(-\partial_z H_y, 0, \partial_x H_y)^{\mathrm{T}}$, which explicitly depends on the spatially periodic permittivity $\epsilon(\mathbf{r})$. To match the boundary conditions, we expand the inverse permittivity as $\frac{1}{\epsilon(\mathbf{r})} = \sum_{m,n=-\infty}^{\infty} \tilde{\epsilon}_{m,n}^{(-1)}e^{imGx}e^{inGy}$, where $\tilde{\epsilon}_{m,n}^{(-1)}$ denotes the Fourier coefficients. By applying the boundary conditions at $z = \pm h/2$, we obtain
\begin{equation}
    \mathrm{det}
    \left[
    \begin{array}{cc}
        g_{0;0} & v_{-1;0}g_{-1;0}\delta \\
        v_{0;-1}g_{0;-1}\delta & g_{-1;-1}
    \end{array}
    \right] = 0,
    \label{eq: 2D guided-mode resonance condition}
\end{equation}
where the subscript $n=0$ and $\sigma=2$ in $v_{mn,\sigma;m'n,\sigma}$ and $g_{mn,\sigma;m'n,\sigma}$ are omitted for simplicity, yielding $ v_{m;m'} $ and $ g_{m;m'} $, respectively. The matrix elements $g_{m;m'}$ are defined as
\begin{equation*}
    g_{m;m'} = 
    \left\{
    \begin{array}{ll}
        Z_{\mathrm{PhC},m} - Z_{\mathrm{b},m'} & \mathrm{if } \; m = m', \\
        \frac{\tilde{\epsilon}_{00}^{(-1)} + \tilde{\epsilon}_{0,1}^{(-1)}/(v_{m;m'}\delta)}{\tilde{\epsilon}_{00}^{(-1)} + \tilde{\epsilon}_{0,1}^{(-1)}v_{m;m'}\delta} & \mathrm{if } \; m \neq m'.
    \end{array}
    \right.
\end{equation*}
The surface impedance of the $m$-th eigenstate, $Z_m = E_{m,\parallel} / H_{m,\parallel}$, is defined in both the background medium and bulk PhC as
\begin{equation*}
    \begin{array}{l}
        Z_{\mathrm{b},m} = ik_{z,m}, \\
        Z_{\mathrm{PhC},0} = (1/\overline{\epsilon} + \tilde{\epsilon}_{0,1}^{(-1)}v_{-1;0}\delta)\partial_z H_{y,0}\Big/H_{y,0}, \\
        Z_{\mathrm{PhC},-1} = (1/\overline{\epsilon} + \tilde{\epsilon}_{0,-1}^{(-1)}v_{0;-1}\delta)\partial_z H_{y,-1}\Big/H_{y,-1},
    \end{array}
\end{equation*}
where $\overline{\epsilon}$ denotes the average permittivity over a unit cell.

The behaviour of the two-band model for the 2D PhC slab is analogous to that of the 1D PhC slab, albeit with increased complexity. In the 2D case, the surface impedance is employed in place of admittance owing to the use of the \textbf{H} field, contrasting with the \textbf{E} field used in the 1D case. The function $g_{mm'}$ is introduced to perform a similar role as that of $f_{mn}$ defined in equation~(\ref{eq:f_mn}) for the 1D scenario. Owing to the periodicity in both the $x$- and $y$-directions in the 2D case, the Fourier components of the permittivity are indexed by two integers, resulting in a more intricate expression for the surface impedance. The vanishing of specific diagonal terms, such as $g_{00} = 0$ or $g_{{-1};{-1}} = 0$, retains its correspondence to the FP-mode and waveguide-mode conditions, respectively.

Following the two-step perturbative approach, we next consider the perturbation near the waveguide mode characterised by $(q{-}G, 0; \omega_0)$. Noting that $g_{{-1};{-1}}(q{-}G,\, 0, \omega_0) = 0$, the guided-mode resonance condition in equation~(\ref{eq: 2D guided-mode resonance condition}) simplifies to:
\begin{equation}
    \mathrm{det}
    \left[
    \begin{array}{cc}
        g_{00} & v_{{-1};0}g_{{-1};0}\delta \\
        v_{0;{-1}}g_{0;{-1}}\delta & \delta\omega\,\partial_{\omega}g_{{-1};{-1}}
    \end{array}
    \right] = 0.
\end{equation}
The leading-order correction to the dispersion is then given by
\begin{equation}
    \delta\omega = -C(\textbf{k}_{||})\,\delta^2,
    \label{eq:C_q in 2D}
\end{equation}
which is of the order $\mathrm{O}(\delta^2)$. The proportionality constant $C(\textbf{k}_{||})$ governs the imaginary part of the resonant mode, analogous to equation~(\ref{eq:C_q}) for the 1D case.

\begin{figure}[h]
    \centering    
    \includegraphics[width=0.75\textwidth]{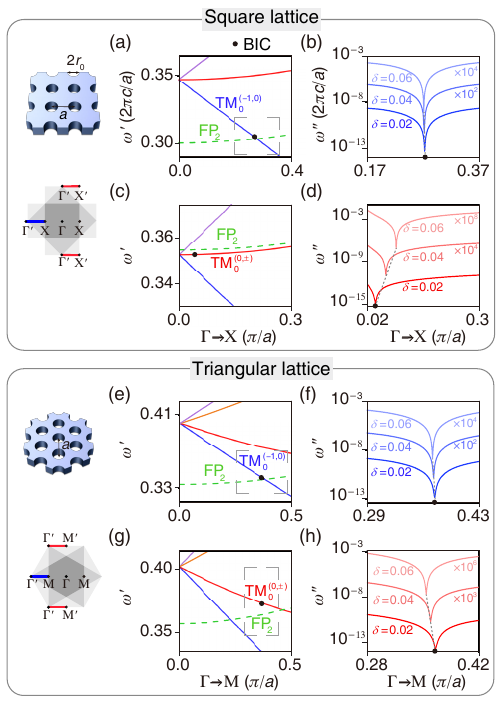}
    \caption{Complex bands and accidental BICs in 2D PhC slabs with square (a--d) and triangular (e--h) lattices. Left insets: schematics of the structure and extended BZ. Primed symbols (e.g. $\Gamma'$, X$'$) denote high-symmetry points that can be folded into the first BZ via reciprocal lattice translations. The folded band diagrams show interactions between the $\mathrm{FP}_2$ band and $\mathrm{TM}_0^{(-1,0)}$ in (a, e), and $\mathrm{TM}_0^{(0,\pm)}$ in (c, g), each giving rise to accidental BICs. (b, f) and (d, h) show the plots of $\omega''$ for TM$_0^{(-1,0)}$ and TM$_0^{(0,\pm)}$, respectively, at the perturbation strengths $\delta = 0.02$, 0.04, and 0.06. As $\delta \to 0$, the BICs converge to fixed points. Slab thicknesses from top to bottom: $h = 1.67a$, $1.41a$, $1.50a$, and $1.40a$; the radius is fixed at $r_0 = a/4$.
}
    \label{fig:2D} 
\end{figure}

The inset of the upper panel of figure~\ref{fig:2D} illustrates an example: a square lattice consisting of cylinders embedded in a background medium. During the perturbation process, the spatial average of the permittivity, $\overline{\epsilon}$, is held constant. The perturbation, with strength $\delta$ singled out, is defined such that $\varepsilon(\mathbf{r}_\parallel) = -1$ outside the cylinders and $\varepsilon(\mathbf{r}_\parallel) = (a^2 - \pi r_0^2)/(\pi r_0^2)$ inside, where $a$ is the lattice period and $r_0$ is the cylinder radius. The corresponding reciprocal lattice of the 2D square lattice in the extended zone scheme is also shown. For example, the equivalent $\Gamma'$ points in the second BZ are located at $\mathbf{k} = (\pm 2\pi/a, 0)$ and $\mathbf{k} = (0, \pm 2\pi/a)$, respectively. The folded bands, shown in figure~\ref{fig:2D}(a), include only those possessing the mirror symmetry $\sigma_v = 1$. The polarisation index $\sigma$ is replaced by the more concrete label `TE' or `TM'. These bands can interact with one another, consistent with the preceding analysis of the mirror symmetry $\sigma_v$ characterising the perturbed bulk Bloch waves that dominate them. Figure~\ref{fig:2D}(b) presents the imaginary part of the TM$_0^{({-1},0)}$ band under variation in the perturbation strength $\delta$. As $\delta \rightarrow 0$, the zeroes of $\omega^{\prime\prime}$ approach the fixed point predicted theoretically, corresponding to $\mathrm{Im}(C) = 0$ in equation~(\ref{eq:C_q in 2D}).

In contrast to the 1D case, the bands in the 2D PhC slabs can also be folded along directions other than $x$, such as in the $y$-direction in square lattices. Because these folded bands may lie above the light line, their interaction with the FP modes should also be considered. However, owing to the degeneracy of the corresponding basis states, for example, $|0, \pm 1; \sigma{=}1,2 \rangle$, the existence of accidental BICs and the minimal dimension of the required Hilbert space warrant further investigation. When the perturbation $\delta \neq 0$ is introduced, this degeneracy is lifted at the order O$(\delta)$ (see supplementary section 5), resulting in the perturbed bulk Bloch waves $\psi_{0,\pm; \sigma, \sigma_v}$. Bloch waves with a different mirror symmetry $\sigma_v$ do not couple; coupling occurs only between states that share the same $\sigma_v$ but differ in the polarisation index $\sigma$. In PhC slabs, the index $\sigma$ corresponds precisely to TE-like or TM-like polarisation along the $k_x$-direction~\cite{sako05}. These two polarisations exhibit significant energy splitting owing to their distinct reflection behaviours at the interface $z = \pm h/2$. Within the framework of the perturbation theory, the interactions between states with smaller energy separations play a dominant role. Therefore, for a given guided-mode resonance band, a two-band model remains sufficient to capture the essential physics, as couplings to more remote bands can be neglected.

As shown in figure~\ref{fig:2D}(c), we consider the interaction between the FP$_2$ and TM$_0^{(0,\pm)}$ bands. A two-band model is adopted, involving the bulk Bloch waves $|\psi_{0, 0; \sigma{=}2} \rangle$ and $|\psi_{0, \pm; \sigma{=}1} \rangle$, which share the same mirror symmetry $\sigma_{v}{=}1$. Importantly, this scenario requires the use of the perturbation theory in the degenerate case of bulk Bloch waves, as the state $|0, \pm; \sigma{=}1\rangle$ is constructed as a linear combination of two plane waves: $|0, \pm; \sigma{=}1\rangle {=} \frac{1}{\sqrt{2}} \left( |0, {-}1; \sigma{=}1 \rangle {-} |0, 1; \sigma{=}1 \rangle \right)$. Following the two-step approach to the perturbation theory, a $2 \times 2$ matrix is derived. The imaginary part $\omega''$ of the TM$_0^{(0,\pm)}$ band can then be obtained from the zeroes of the determinant of this matrix, as shown in figure~\ref{fig:2D}(d). However, it is observed that the wavevector of the accidental BIC shifts significantly with variation in $\delta$, in contrast to the more stable behaviour seen in figure~\ref{fig:2D}(b).

The results presented above demonstrate that bands folded along the $y$-direction are analogous to those folded along the $x$-direction in their capability to give rise to accidental BICs via their interaction with FP modes. The primary distinction lies in the degeneracy of the associated basis states. For modes along the $k_x$-axis, the bands folded from the $y$-direction with wavevector shifts $\pm G$ are twofold degenerate; therefore, analogous to the equation~(\ref{eq:eigen-sys-pertur-degeneracy}) in the 1D case, the correction to the real part $\omega'$ scales as O$(\delta)$. The detailed derivation of the eigenvalue correction in the 2D case is provided in supplementary section 5. In contrast, for the bands folded along the $x$-axis, the leading-order correction to $\omega'$ is O$(\delta^2)$, as described in equation~(\ref{eq:C_q in 2D}). As a result, the accidental BICs arising from $y$-folded bands exhibit a more pronounced shift in the response to perturbation.

In addition to the square lattice, the theoretical framework described above can be readily extended to other lattice geometries, such as a triangular lattice. This extension requires only a change in the periodic permittivity modulation $\varepsilon(\mathbf{r}_\parallel)$ to reflect the new geometry. Specifically, the perturbation is defined as $\varepsilon(\mathbf{r}_\parallel) = -1$ outside the cylinders and $\varepsilon(\mathbf{r}_\parallel) = (\sqrt{3}a^2/2 - \pi r_0^2)/(\pi r_0^2)$ inside, ensuring that the spatial average $\langle \varepsilon(\mathbf{r}_\parallel) \rangle = 0$. We focus on modes along the high-symmetry directions, where the minimal Hilbert space giving rise to a nonzero $\omega^{\prime\prime}$ remains two-dimensional. For instance, along the $\Gamma$–M direction, as shown in the inset of the lower panel of figure~\ref{fig:2D}, the guided-mode resonance condition continues to be governed by equation~(\ref{eq: 2D guided-mode resonance condition}). Figure~\ref{fig:2D}(e) and (f) show the folded bands of the waveguide and FP modes in the triangular lattice, along with the corresponding imaginary parts of the TM$_0^{(-1,0)}$ band under varying $\delta$. As $\delta \rightarrow 0$, the condition $\omega^{\prime\prime} = 0$ again converges to the theoretical prediction $\mathrm{Im}(C) = 0$. Moreover, as shown in figure~\ref{fig:2D}(g) and (h), accidental BICs also emerge when the FP modes interact with bands folded along other directions with wavevector shifts $(-1, \pm\sqrt{3})G/2$ in triangular lattices. This observation further confirms the generality of the formation mechanism for accidental BICs. All accidental BICs discussed above satisfy the scaling law $Q \propto 1/(q-q_*)^2$ , as verified numerically; see supplementary section 2 for detailed verification.

The two examples discussed above, namely, square and triangular lattices, demonstrate that the two-band model, consisting of an FP mode and a guided-mode resonance, can effectively capture the origin of nonzero $\omega^{\prime\prime}$ and predict the existence of `accidental' BICs in 2D PhC slabs. Furthermore, band folding along the $x$-direction can lead to crossings between different branches of guided-mode resonances, such as the TM$_0^{(1,0)}$ and TM$_2^{(-1,0)}$ modes, resulting in the interaction between three bands, analogous to the 1D case. However, in 2D, the periodicity defined by the lattice vectors $\mathbf{a}_1$ and $\mathbf{a}_2$ corresponds to two reciprocal lattice vectors, $\mathbf{G}_1$ and $\mathbf{G}_2$, respectively, which give rise to additional types of crossing points upon band folding. These crossings can occur between different branches of guided-mode resonances, corresponding to the folding of linear combinations of $\mathbf{G}_1$ and $\mathbf{G}_2$. Near such crossing points, the two-band model becomes insufficient to capture the underlying physics; instead, a minimal Hilbert space of three dimensions is required. We analyse these crossing points separately below.

\begin{figure}[h]
    \centering    
    \includegraphics[width=0.8\textwidth]{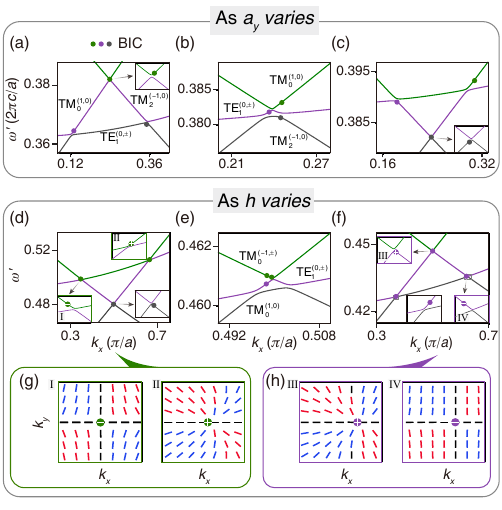}
    \caption{Evolution of Friedrich--Wintgen BICs in 2D PhC slabs. (a) Three guided-mode resonance bands—TM$_0^{(1,0)}$, TM$_2^{(-1,0)}$ (folded along the $x$-direction), and TE$_1^{(0,\pm)}$ (folded along the $y$-direction)—interact pairwise to form three BICs. As the period in the $y$-direction decreases from $a_y = a$ in (a) to $0.95a$ in (b) and $0.925a$ in (c), the TE$_1^{(0,\pm)}$ band blue-shifts, whereas the BICs remain on their respective bands. (d--f) Interaction between TM$_0^{(1,0)}$, TE$_1^{(0,\pm)}$, TM$_0^{(-1,\pm)}$: two BICs emerge on the upper band in (d), then merge, annihilate, and re-emerge as $h$ increases from 0.67$a$ to 0.81$a$ and $a$. Insets show the details near the crossing points. (g, h) Far field polarisation for (g) $h=0.67a$, (h) $h=a$ show opposite topological charges. Parameters: $a_x=a$, $h=2a$, $r_0=a/4$, $\delta=0.1$ in (a--c); $a_x=a_y=a$, $\delta=0.01$ in (d--f).
}
    \label{fig:3bandBIC-2D}
\end{figure}

Let us take the square lattice as an example. The first type of crossing is analogous to that between two guided-mode resonance bands in the 1D system, as discussed in section~\ref{sec:FW-BIC}. For instance, as shown in figure~\ref{fig:3bandBIC-2D}(a), we focus on the interaction between the TM$_0^{(1,0)}$ and TM$_2^{(-1,0)}$ bands. Near this crossing point (highlighted by the dashed box), a three-band model captures the interaction between these two guided-mode resonances and an FP mode with the same $\sigma_{v}$ symmetry. In fact, the above three modes possess an eigenvalue of $+1$ under $\sigma_v$ reflection, corresponding to an even symmetry of $E_z$ with respect to the $x$-$z$ plane. The relevant three Bloch waves are $|\psi_{0,0;\sigma{=}2}\rangle$, $|\psi_{{-1},0;\sigma{=}2}\rangle$, and $|\psi_{1,0;\sigma{=}2}\rangle$ with mirror symmetry $\sigma_v=1$. The perturbed bulk Bloch wavefunctions, as shown in equation~(\ref{eq: perturbed lambda for 2D}), can be rewritten as
\begin{equation*}
    \begin{array}{l}
        |\psi_{00;2}\rangle = |00;2\rangle + v_{00,2;{-1},0;2}|{-1},0;2\rangle\,\delta + v_{00,2;1,0,2}|1,0;2\rangle\,\delta + \cdots, \\
        |\psi_{{-1},0;2}\rangle = |{-1},0;2\rangle + v_{{-1},0,2;00,2}|00;2\rangle\,\delta + v_{{-1},0,2;10,2}|1,0;2\rangle\,\delta + \cdots, \\
        |\psi_{1,0;2}\rangle = |1,0;2\rangle + v_{1,0,2;00,2}|00;2\rangle\,\delta + v_{1,0,2;{-1},0,2}|{-1},0;2\rangle\,\delta + \cdots. \\
    \end{array}
\end{equation*}
By matching the boundary conditions at the interface $z = \pm h/2$, we derive a set of three linear equations, leading to a $3 \times 3$ matrix. The zeroes of its determinant yield the complex resonant frequencies $\omega - i\omega''$ of the guided-mode resonances and FP modes. A BIC is readily identified as the point where the resonant frequency becomes purely real.

Another type of crossing point arises from the interaction between bands folded along the $x$- and $y$-directions~\cite{Zhou2025tailoring}, necessitating an analysis of the minimal dimension of the Hilbert space. The primary distinction lies in the degeneracy of the basis states associated with the wavevector shifts of $\pm G$ along the $y$-direction. As demonstrated in the analysis of the accidental BIC formed via the interaction between FP modes and $y$-folded bands, this degeneracy does not increase the required dimension of the Hilbert space, and a two-band model is sufficient to describe the interaction. By performing a similar analysis, a three-band model is also sufficient to capture the essential physics near the crossing point of two guided-mode resonance bands.

For example, near the crossing point between the $\mathrm{TM}_2^{(-1,0)}$ and $\mathrm{TE}_1^{(0,\pm)}$ bands shown in figure~\ref{fig:3bandBIC-2D}(a), the interaction can be described by a three-band model involving the Bloch waves $|\psi_{0,0;\sigma{=}2}\rangle$, $|\psi_{{-1},0;\sigma{=}2}\rangle$, and $|\psi_{0,\pm;\sigma{=}2}\rangle$. The mirror symmetry label $\sigma_v = 1$ is omitted. Using the aforementioned two-step perturbation approach, the perturbed bulk Bloch wavefunctions are expressed as
\begin{equation*}
    \begin{array}{l}
        |\psi_{00;2}\rangle = |00;2\rangle + v_{00,2;-1,0,2}|{-}1,0;2\rangle\,\delta + v_{00,2;0,\pm,2}|0,\pm;2\rangle\,\delta + \cdots, \\
        |\psi_{-1,0;2}\rangle = |{-}1,0;2\rangle + v_{-1,0,2;00,2}|00;2\rangle\,\delta + v_{-1,0,2;0,\pm,2}|0,\pm;2\rangle\,\delta + \cdots, \\
        |\psi_{0,\pm;2}\rangle = |0,\pm;2\rangle + v_{0,\pm,2;00,2}|00;2\rangle\,\delta + v_{0,\pm,2;-1,0,2}|{-}1,0;2\rangle\,\delta + \cdots. \\
    \end{array}
\end{equation*}
Here, the state $|0,\pm;\sigma{=}2\rangle {=} \frac{1}{\sqrt{2}}(|0,{-}1;\sigma{=}2\rangle {+} |0,1;\sigma{=}2\rangle)$ represents a linear combination of two plane waves. A $3 \times 3$ matrix can be constructed by matching the boundary conditions at the interface $z = \pm h/2$. The real and imaginary parts of the resonant frequencies, as well as the positions of the BICs, can then be determined from the vanishing of the determinant of this matrix.

The interaction between the $\mathrm{TM}_0^{(1,0)}$ and $\mathrm{TE}_1^{(0,\pm)}$ bands can be treated analogously, as also shown in figure~\ref{fig:3bandBIC-2D}(a). In this case, the basis state $|{-1},0;\sigma{=}2\rangle$ is replaced by $|1,0;\sigma{=}2\rangle$. Thus, the three relevant Bloch states become $|\psi_{00;\sigma{=}2}\rangle$, $|\psi_{1,0;\sigma{=}2}\rangle$, and $|\psi_{0,\pm;\sigma{=}2}\rangle$. The position of the Friedrich--Wintgen BIC can again be determined from the poles of the $S$-matrix.

As illustrated in figure~\ref{fig:3bandBIC-2D}(a), the three bands $\mathrm{TM}_0^{(1,0)}$, $\mathrm{TM}_2^{(-1,0)}$, and $\mathrm{TE}_1^{(0,\pm)}$ interact pairwise, giving rise to three BICs near their respective crossing points. Notably, the positions of these BICs shift in response to changes in the system parameters. For instance, decreasing the period in the $y$-direction causes a blue-shift of the $\mathrm{TE}_1^{(0,\pm)}$ band, as shown in figure~\ref{fig:3bandBIC-2D}(b) and (c). These three BICs are particularly robust, as they are distributed across three different bands and no gap closes during their evolution. In the vicinity of the crossing points in figure~\ref{fig:3bandBIC-2D}(a) and (c), the three-band model remains valid for capturing the band interactions and predicting the BIC positions. In contrast, figure~\ref{fig:3bandBIC-2D}(b) illustrates an `accidental' degeneracy, where three guided-mode resonances intersect. Near this degenerate point, a four-band model is required, incorporating an FP mode. The corresponding Bloch wavefunctions include $|\psi_{00;\sigma{=}2}\rangle$, $|\psi_{-1,0;\sigma{=}2}\rangle$, $|\psi_{1,0;\sigma{=}2}\rangle$, and $|\psi_{0,\pm;\sigma{=}2}\rangle$.

The annihilation and generation of Friedrich--Wintgen BICs in the 2D PhC slab can be described using this four-band model. As shown in figure~\ref{fig:3bandBIC-2D}(d), the bands $\mathrm{TM}_0^{(1,0)}$, $\mathrm{TE}_1^{(0,\pm)}$, and $\mathrm{TM}_0^{(-1,\pm)}$ interact, resulting in three BICs, two of which are located on the upper band. When a structural parameter, such as the thickness $h$, is varied, these two BICs undergo merging, annihilation, and regeneration. The far-field polarisation states near these two BICs are computed at specific thicknesses $h = 0.67a$ and $h = a$, revealing opposite topological charges, in accordance with charge conservation during the evolution. We further examine the polarisation states near Friedrich--Wintgen BICs under varying perturbation strengths, with the vortex position shifting along the dispersion curve while maintaining its charge (see supplementary section 4).

Similar to the case in figure~\ref{fig:3bandBIC-2D}(b), figure~\ref{fig:3bandBIC-2D}(e) shows another critical point corresponding to an accidental degeneracy of three guided-mode resonances, necessitating a four-band model to fully describe the interactions. Furthermore, to obtain the far-field polarisation of resonant states away from the high-symmetry line, both polarisations $\sigma = 1,2$ should be included, thereby doubling the number of relevant Bloch states. Consequently, the minimal Hilbert space has a dimensionality of 6 for the three-band model, or 8 for the four-band model, when considering the far-field polarisation off the high-symmetry line.

\section{Summary}

In this study, a systematic investigation of the complex band structure of PhC slabs was conducted from a first-principles perspective. The complex band structure was rigorously defined in terms of the poles of the scattering matrix, and the behaviour of these poles was elucidated using perturbation theory within a minimally constructed Hilbert space. The analysis revealed that the minimal dimension of this Hilbert space, or equivalently the number of scattering channels in the scattering matrix, is determined by the number of bulk Bloch waves participating in the resonant modes.

In addition to predicting the real parts of the complex bands, that is, the dispersion relations, the imaginary parts were accurately determined, enabling a quantitative description of the resonant modes. It was shown that the imaginary part scales quadratically with the perturbation strength $\delta$, with a proportionality coefficient $C$ that depends on the lattice type, slab thickness, wavevector, and frequency, analogous to the structure factor. All known types of BICs, including accidental, Friedrich--Wintgen, and symmetry-protected BICs, can be identified within this first-principles framework. Other physical properties, such as far-field polarisation and band singularities, are also captured. The main results are summarised below, categorised according to the required number of scattering channels, which also corresponds to the number of bands considered in each model.

\textbf{Waveguide and FP modes (one scattering channel):} The minimal $S$-matrix required to describe their dispersion has $\mathrm{dim}(S)=1$. For waveguide modes lying below the light line, the `outgoing' waves are evanescent, and the field profiles are dominated by the corresponding bulk Bloch wave $\psi_{q - nG}$, where $n$ denotes the band-folding index. By contrast, FP modes exhibit complex frequencies owing to radiative losses and are governed by the Bloch wave $\psi_q$, with $q$ lying inside the light cone, which accounts for their leaky character.

\textbf{Guided-mode resonances and accidental BICs (two scattering channels):} In this case, a $2\times2$ $S$-matrix is required. By constructing the $S$-matrix from two Bloch waves ($\psi_q$ and $\psi_{q-nG}$), both the real and imaginary parts of the guided-mode resonances can be accurately determined. The imaginary part again scales quadratically with $\delta$. The condition for accidental BICs is derived and shown to correspond to a twofold degeneracy in the eigenvalues of the surface impedance matrix. These BICs appear as fixed points of the perturbation and are independent of $\delta$. The degeneracy of the impedance matrix further reveals the presence of dual solutions, characterised by the vanishing of the $n$-th diffraction order outside the slab. Notably, the `accidental' BIC arises within the guided-mode resonance band, whereas its dual appears in the FP band. Together, these results provide a comprehensive physical picture of complex band formation.

\textbf{Friedrich--Wintgen and symmetry-protected BICs (three scattering channels):} Here, a $3\times3$ $S$-matrix is required. The interaction between two guided-mode resonances typically leads to an avoided crossing, with a bandgap proportional to $\delta$. Friedrich--Wintgen BICs emerge at the crossing points of these bands and shift linearly with increasing $\delta$. In addition, a dual solution, termed the `dual BIC', is identified; it emerges in a neighbouring guided-mode band but is destroyed by mode-mixing effects. Symmetry-protected BICs and the nearby high-$Q$ resonant states are also analysed. By examining the proportionality constant $\tilde{C}$ in $\delta\omega'' = \tilde{C} \cdot (q - q_{\mathrm{BIC}})^2 \cdot \delta^2$, criteria for maintaining high-$Q$ factors over broad wavevector ranges are established. These insights provide practical guidance for the design of robust high-$Q$ resonances.

\textbf{Far-field polarisation states (four scattering channels) and EPs (six channels):} To account for polarisation degrees of freedom in far-field radiation, a $4\times4$ $S$-matrix is required, effectively doubling the number of Bloch states. This expanded basis enables the calculation of complex bands across the full BZ. Far-field polarisation states and polarisation singularities, characterised by nonzero winding numbers, are identified in momentum space. When an FP mode and two guided-mode resonances with orthogonal polarisations are included, a $6\times6$ $S$-matrix becomes necessary. Within this framework, EPs are identified and their evolution is analysed. Specifically, EPs emerge at the crossings of orthogonally polarised resonances along high-symmetry directions and shift off-axis in proportion to $\delta$.

\textbf{Extension to 2D PhC slabs:} In comparison with 1D systems, 2D PhC slabs exhibit more complex band folding in the reciprocal space. Nevertheless, the minimal scattering channels necessary to describe nonzero $\omega''$ for guided-mode resonances and BICs can still be determined analogously. Using square and triangular lattices as representative examples, we show that accidental BICs formed by the interaction between a guided-mode resonance and an FP mode require two scattering channels. In contrast, the interactions between two guided-mode resonances, especially those corresponding to the folding by linear combinations of reciprocal lattice vectors, produce Friedrich--Wintgen BICs near the band-crossing points, requiring three scattering channels. By employing the scattering matrix formalism, we analyse the evolution of guided-mode resonance bands under geometric tuning, revealing merging, annihilation, and regeneration of Friedrich--Wintgen BICs, thus underscoring the rich topological structure of complex bands in 2D systems.

This study advances the understanding of complex band structure in PhC slabs through the analysis of the poles of the scattering matrix. It provides a unified explanation for all known BIC types and enhances our comprehension of light confinement in periodic media. Nevertheless, the present framework is based on perturbation theory and is therefore valid in the perturbation regime, where only leading-order contributions to the scattering matrix are retained. For stronger modulations or higher-index-contrast structures \cite{chang12}, higher-order terms may become important and modify predictions on the BIC positions, band gaps near avoided crossing points, and $Q$ factors of resonance modes, particularly on the phenomena such as the merging and reappearance of BICs and the closing and reopening of bang gaps under large-scale parameter tuning. In such regime, extending the scattering matrix treatment to include higher-order contributions would be necessary. However, the framework itself is not inherently limited to weak perturbations and can, in principle, be generalized by incorporating higher-order terms, allowing it to be extended to high-index-contrast PhC slabs and strongly modulated metasurfaces.

These findings suggest several promising future directions. One is the design of ultra-high-$Q$ resonances by exploiting the dependence of the $Q$ factor on the proportionality coefficient $\tilde{C}$, potentially via automated optimisation for applications in sensing, lasing, and nonlinear optics. Additionally, the framework of the perturbation theory is inherently well suited for introducing symmetry-breaking perturbations, enabling a rigorous investigation of the polarisation dynamics, particularly the generation and evolution of circularly polarised points and related phenomena. This framework facilitates applications requiring the tailoring of optical resonances and chiral polarisation states. Moreover, the scattering matrix formalism provides a new perspective on the emergence and evolution of EPs in open periodic systems and offers a refined approach for analysing the interplay between EPs and optical vortex states such as BICs and circularly polarised points. The perturbation theory and scattering matrix formalism established here can also be generalised to other systems—acoustic, electronic, and beyond—enabling broad investigations into complex band structure across diverse physical platforms.

\section*{Data availability statement}
All data supporting the findings of this study are provided here: https://doi.org/10.6084\\
/m9.figshare.29551967. Additional data and related codes are available upon reasonable request from the authors.

\section*{Acknowledgments}
D.H. gratefully thanks Prof. C. M. Song and Dr. Jun Wang for valuable discussions. This work was supported by the National Natural Science Foundation of China (grant nos. 12574409, 12204388, 12547101).

\section*{Author contributions}
D.H. conceived the ideas and designed the project. C.Z., and D.H. developed the theory. J.L., Z.P. and Q.S. performed the theoretical and numerical calculations. Q.S. and D.H. interpreted the results and co-wrote the manuscript. All authors contributed to analysis and discussion of the results.

\section*{Conflict of interest}
The authors declare no competing interests.

\section*{Reference}
\bibliography{reference}

\end{document}